\documentclass[fleqn,usenatbib]{mnras}

\usepackage[T1]{fontenc}

\DeclareRobustCommand{\VAN}[3]{#2}
\let\VANthebibliography\thebibliography
\def\thebibliography{\DeclareRobustCommand{\VAN}[3]{##3}\VANthebibliography}

\usepackage{graphicx}	
\usepackage{amsmath}	
\usepackage{soul}
\usepackage{amssymb}	
\usepackage{newtxtext,newtxmath}
\usepackage{natbib}

\newcommand{\Rgal}{$R_{\rm 26.5}$}
\newcommand{\Rhalf}{$R_{\rm 50}^{\star}$}
\newcommand{\Rt}{$R_{\rm 90}^{\star}$}
\newcommand{\deltar}{$\Delta r_{\rm DM}$}
\newcommand{\Rcrit}{$R_{\rm 200}$}
\newcommand{\Mcrit}{$M_{\rm 200}$}
\newcommand{\Am}{$A_{\rm 1}$}

\newcommand{\mus}{$\mu_\star$}
\newcommand{\Mhalf}{$M_{50}^{\star}$}
\newcommand{\Mt}{$M_{90}^{\star}$}

\title[Lopsided galaxies in a cosmological context]{Lopsided Galaxies in a cosmological context: a new galaxy-halo connection}

\author[Varela-Lavin et al.]{
Silvio Varela-Lavin$^{1,2}$\thanks{E-mail: silvio.varela@userena.cl},
Facundo A. G\'omez$^{1,2}$,
Patricia B. Tissera$^{3,4}$,
Gurtina Besla$^{5}$,\newauthor
Nicolás Garavito-Camargo$^{6}$,
Federico Marinacci$^{7}$, and
Chervin F. P. Laporte$^{8,9,10}$
\\
$^{1}$Departamento de F\'isica y Astronom\'ia, Universidad de La Serena, Av. Juan Cisternas 1200 Norte, La Serena, Chile.\\
$^{2}$ Instituto de Investigaci\'on Multidisciplinar en Ciencia y Tecnolog\'ia, Universidad de La Serena, Ra\'ul Bitr\'an 1305, La Serena, Chile.\\
$^{3}$Instituto de Astrof\'{i}sica, Pontificia Universidad Cat\'olica de Chile, Av. Vicuña Mackenna 4860, Santiago, Chile.\\
$^{4}$Centro de Astro-Ingenier\'ia, Pontificia Universidad Cat\'olica de Chile, Av. Vicu\~na Mackenna 4860, Santiago, Chile.\\
$^{5}$Steward Observatory, University of Arizona, 933 North Cherry Avenue,Tucson, AZ 85721, USA.\\
$^{6}$Center for Computational Astrophysics, Flatiron Institute, 162 Fifth Avenue, New York, NY 10010, USA\\
$^{7}$Department of Physics and Astronomy "Augusto Righi" University of Bologna via Gobetti 93/2 40129 Bologna, Italy\\
$^{8}$Departament de Física Quàntica i Astrofísica (FQA), Universitat de Barcelona (UB), c. Martí i Franquès, 1, 08028 Barcelona, Spain e-mail: tantoja@fqa.ub.edu\\
$^{9}$Institut de Ciències del Cosmos (ICCUB), Universitat de Barcelona (UB), c. Martí i Franquès, 1, 08028 Barcelona, Spain\\
$^{10}$Institut d’Estudis Espacials de Catalunya (IEEC), c. Gran Capità, 2-4, 08034 Barcelona, Spain\\
}
\date{Accepted XXX. Received YYY; in original form ZZZ}

\pubyear{2023}

\begin{document}
\label{firstpage}
\pagerange{\pageref{firstpage}--\pageref{lastpage}}
\maketitle

\defcitealias{Reichard2008}{R08}
\begin{abstract}
Disc galaxies commonly show asymmetric features in their morphology, such as warps and lopsidedness. These features can provide key information regarding the recent evolution of a given disc galaxy. In the nearby Universe, up to $\sim30$ percent of late-type galaxies display a global non-axisymmetric lopsided mass distribution. However, the origin of this perturbation is not well understood. In this work, we study the origin of lopsided perturbations in simulated disc galaxies extracted from the TNG50 simulation of the IllustrisTNG project. We statistically explore different excitation mechanisms for this perturbation, such as direct satellite tidal interactions and distortions of the underlying dark matter distributions. We also characterize the main physical conditions that lead to lopsided perturbations. 
50 percent of our sample galaxy have lopsided modes $m=1$ greater than $\sim 0.12$. We find a strong correlation between internal galaxy properties, such as central stellar surface density and disc radial extension with the strength of lopsided modes. The majority of lopsided galaxies have lower central surface densities and more extended discs than symmetric galaxies. As a result, such lopsided galaxies are less self-gravitationally cohesive, and their outer disc region is more susceptible to different types of external perturbations. However, we do not find strong evidence that tidal interactions with satellite galaxies are the main driving agent of lopsided modes. Lopsided galaxies tend to live in  asymmetric dark matter halos with high spin, indicating strong galaxy-halo connections in late-type lopsided galaxies.

\end{abstract}


\begin{keywords}
galaxies: spiral --galaxies: evolution -- galaxies: formation -- galaxies: haloes -- galaxies: structure -- galaxies: interactions
\end{keywords}



\section{Introduction}
\label{sec:intro}

In the nearby Universe spiral galaxies, such as our own, show different morphological asymmetries such as warps, lopsidedness and polar rings, among others. Lopsided perturbations in disc galaxies are one of the most common. It is described as a morphological distortion in which a side of the disc is more elongated than the other. Such global non-axisymmetric perturbation is typically quantified through a Fourier decomposition of the mass or light distribution, focusing on the $m=1$ mode, \Am\  \citep{Rix1995,Quillen2011}. \cite{Rix1995} showed that, for lopsided galaxies, the amplitude of \Am\ increases with radius in the outer galaxy regions. Clear examples of lopsided galaxies include M101 or NGC1637. 

One of the first studies reporting this perturbation was presented by \cite{Baldwin1980}, who analyzed the spatial distribution of HI gas in the outer regions of a sample of galaxies.  Lopsidedness  has been studied in the stellar  \citep{Rix1995} and HI gas distributions \citep{RichterSancisi1994,Haynes1998} of galaxies, as well as  on their large-scale kinematics  \citep{Swaters1999,Schoenmakers1997,Khademi2021}, and compared against numerical models \citep{Ghosh2022,2022A&A...662A..53L}. In the nearby Universe 30 percent of late-type galaxies show high values of \Am \citep{Zaritsky_1997,Bournaud2005}. On the other hand, for early-type galaxies the frequency with which this perturbation arises is close to 20 percent \citep{Rudnick1998}. This  higher frequency of lopsidedness in late-type galaxies was confirmed by \cite{Conselice2000}, who analyzed  a sample of 113 galaxies both early and late-type.  Lopsidedness in this sample was quantified using the 180° rotational asymmetry measure, $\rm A_{180}$. They found a strong relation between morphology and lopsidedness, showing that early-type galaxies (elliptical and lenticular) tend to systematically have lower values of $\rm A_{180}$. 

A more recent study from \citet[][hereafter \citetalias{Reichard2008}] {Reichard2008} measured the asymmetries in galaxies through \Am\ using their surface brightness distribution in three different bands. Their sample consisted of more than 25000 galaxies from Sloan Digital Sky Survey (SDSS). They showed that  the occurrence and strength of lopsidedness has a strong dependency with galaxy structural properties. Disc galaxies with higher \Am\ tend to have low stellar mass, concentration and high central stellar density. The latter is the parameter that most clearly correlates with the lopsidedness. As in \cite{Rix1995}, \citetalias{Reichard2008} shows that the amplitude of the $m=1$ mode is negligible in the very inner regions of galactic discs due to its strong self-gravitating nature. However  a systematic increase of  the \Am\;  parameter with galactocentric radius is observed in the outer galactic regions of lopsided galaxies. In addition \citetalias{Reichard2008} finds that the lopsided light distributions are primarily caused by lopsided distributions in the stellar mass. 

As discussed by \cite{Jog2009}, lopsidedness can have very significant effects on the evolution of galaxies. In particular, for disc galaxies it can induce the redistribution of stellar mass due to angular momentum transport and the modulation  of hosts  star formation histories. In addition, the internal torques induced by such $m=1$ modes can result in the loss of angular momentum by the host gaseous disc, thus affecting the growth of the central supermassive black hole. As a result, lopsided perturbations could allow us to place important constrains on the recent interaction history of galaxies.

Several studies that have tried to characterize the main mechanisms driving lopsided perturbations. Possible proposed mechanism are minor mergers \citep{Walker1996,Zaritsky_1997, Ghosh2022} and tidal interactions due to close encounters between galaxies of similar mass \citep{Kornreich2002}. Indeed, low density galaxies and, in particular the outskirts of galactic disc, are likely to be more susceptible to tidal stress. However, a study of 149 galaxies observed in the near-infrared from the OSUBGS sample \citep{OSUBGS} by \citet{Bournaud2005} found that the amplitude of the $m=1$ mode is uncorrelated with the presence of companions. Instead, they suggested that asymmetric  gas accretion is an important driver of lopsidedness. Similarly, \citet[]{2022A&A...662A..53L} used a sample of simulated galaxies extracted from the TNG100 simulation of the IllustrisTNG project \citep{Nelson2019a} to study the origin of these perturbations. They concluded that the most frequent mechanism for the formation of lopsided discs is asymmetric star formation, probably related to gas accretion. However, they also observed that the distortions in the gas and stars were not strongly correlated.

Another plausible mechanism driving lopsided discs relates to perturbations in the density field of the underlying galactic dark matter (DM) halo. These asymmetries in a DM halo could be produced by a resonant interaction between the DM halo particles and an orbiting satellite. The resulting asymmetry of the DM overdensity field, or wake, can be thought of as a superposition of different modes excited by such resonant interaction. The wake's associated torque, exerted on the embedded disc,  could lead to the formation of strong morphological disturbances such as lopsidedness and warps, among others. Indeed, \citet[][] {Weinberg1998} showed that such perturbations can induce the formation of vertical patterns, such as warps and corrugation patterns. These results were latter confirmed using fully cosmological hydrodynamical simulations \citep{Gomez2016} as well as carefully tailored simulations to study the response of the Milky Way halo to a recently accreted Large Magellanic  Cloud satellite \citep{Laporte2018a,GaravitoCamargo2019}. Furthermore, as discussed by \citet[][]{Jog1999}, these DM halo asymmetries can also induce the formation of lopsided perturbations, and  sustain them for long periods of time. Using the Millennium simulation \citep{Springel2005},  \citet{gao} characterized asymmetries in  DM halos within a mass range of $\sim 10^{12} $ to $10^{15}$ $M_{\odot}$. The asymmetries were quantified based on shifts between the overall DM halo center of mass (CoM) and its center of density (cusp). Shifts between the a system's CoM and cusp can be though as the a dipolar component of a wake \citep{Weinberg1998,GaravitoCamargo2021a}, and typically have the strongest amplitude of all modes. They showed that such asymmetries were not uncommon and that the frequency with which they arose depended on the host mass. While 20 percent of cluster haloes have CoM  separated from their cusp  by  distances larger than 20 per cent of the virial radius, only 7 per cent of the Milky Way-mass haloes show such large asymmetries.

Despite all these studies, several questions remain to be answered regarding lopsided galaxies, including the main driver and longevity of such perturbation. Additionally, we do not yet understand whether lopsidedness can be linked to fundamental properties of the structure and evolution of the host galaxy and its halo. In this work we  analyze a large sample of late-type galaxies, extracted from the  Illustris TNG50  project \citep{Nelson2019a, Pillepich2019MNRAStng50} to shed light on these issues. This highly-resolved fully-cosmological hydrodynamical simulation includes, in a self-consistent manner, the different physical processes that have been proposed as the main drivers of morphological perturbations. In particular, we focus on Milky Way mass-like halos, whose stellar disc can be resolved with the available mass resolution. In Section \ref{sec:simulation_data} we discuss the details of the numerical simulation, as well as the selection criteria for our galaxy sample. The methods to characterize the properties of the stellar discs, and to quantify the presence of a lopsided mode on their density distribution, are introduced in Section \ref{sec:methods}. In Section \ref{sec:results} we present our results. Our conclusion and discussion  are summarised  in Section \ref{sec:conclu}

\begin{figure}
\begin{center}
\includegraphics[trim={0cm 0cm 0cm 0cm},clip,width=\linewidth]{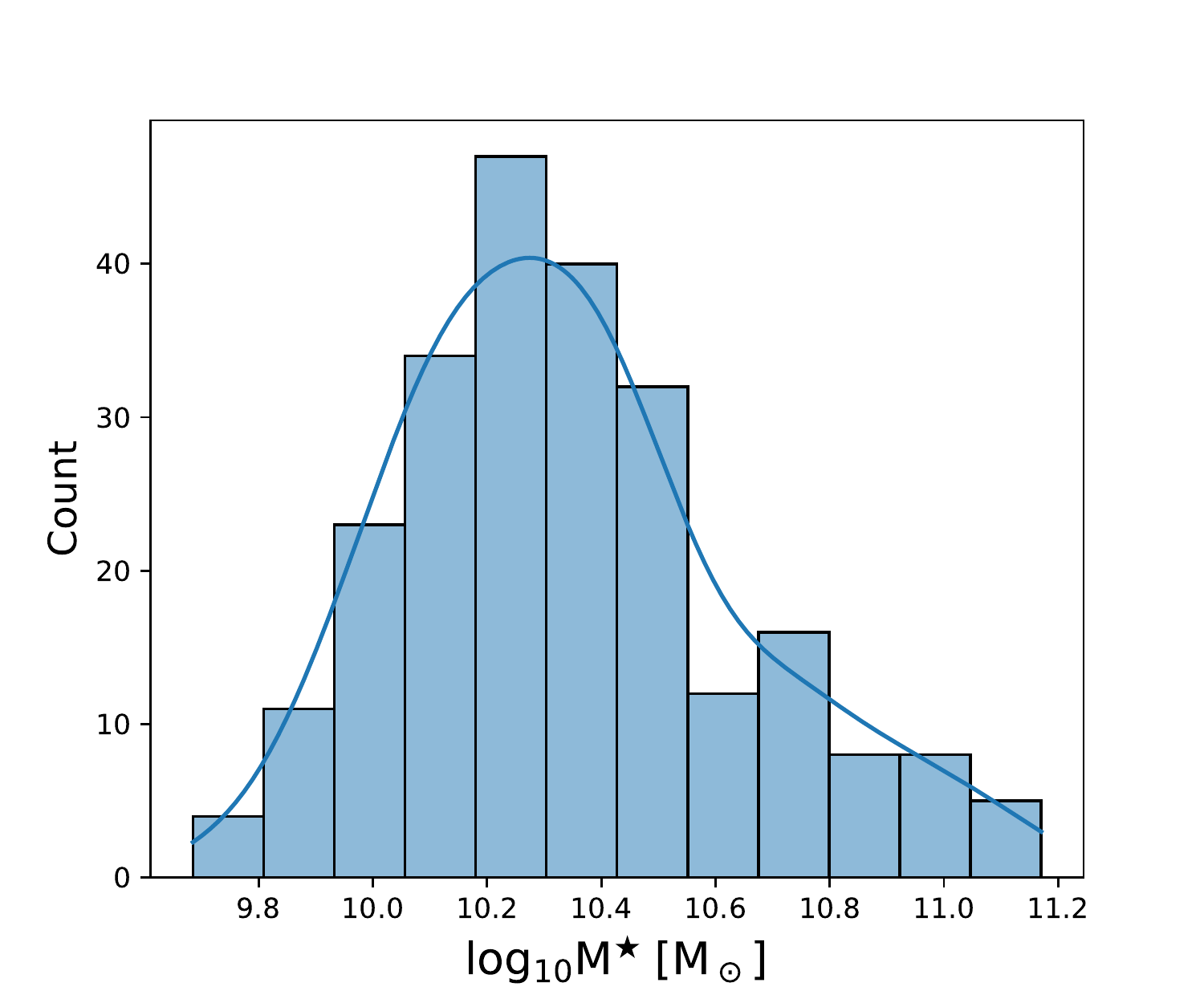}
\caption{Stellar mass distribution of our selected sample (see Section 2.2). 
The solid line depicts a KDE of this distribution. }
\label{statesample_Mstar}
\end{center}
\end{figure}

\begin{figure*}
\begin{center}

\includegraphics[trim={5.5cm 3cm 4.5cm 2.8cm},clip,width=\linewidth]{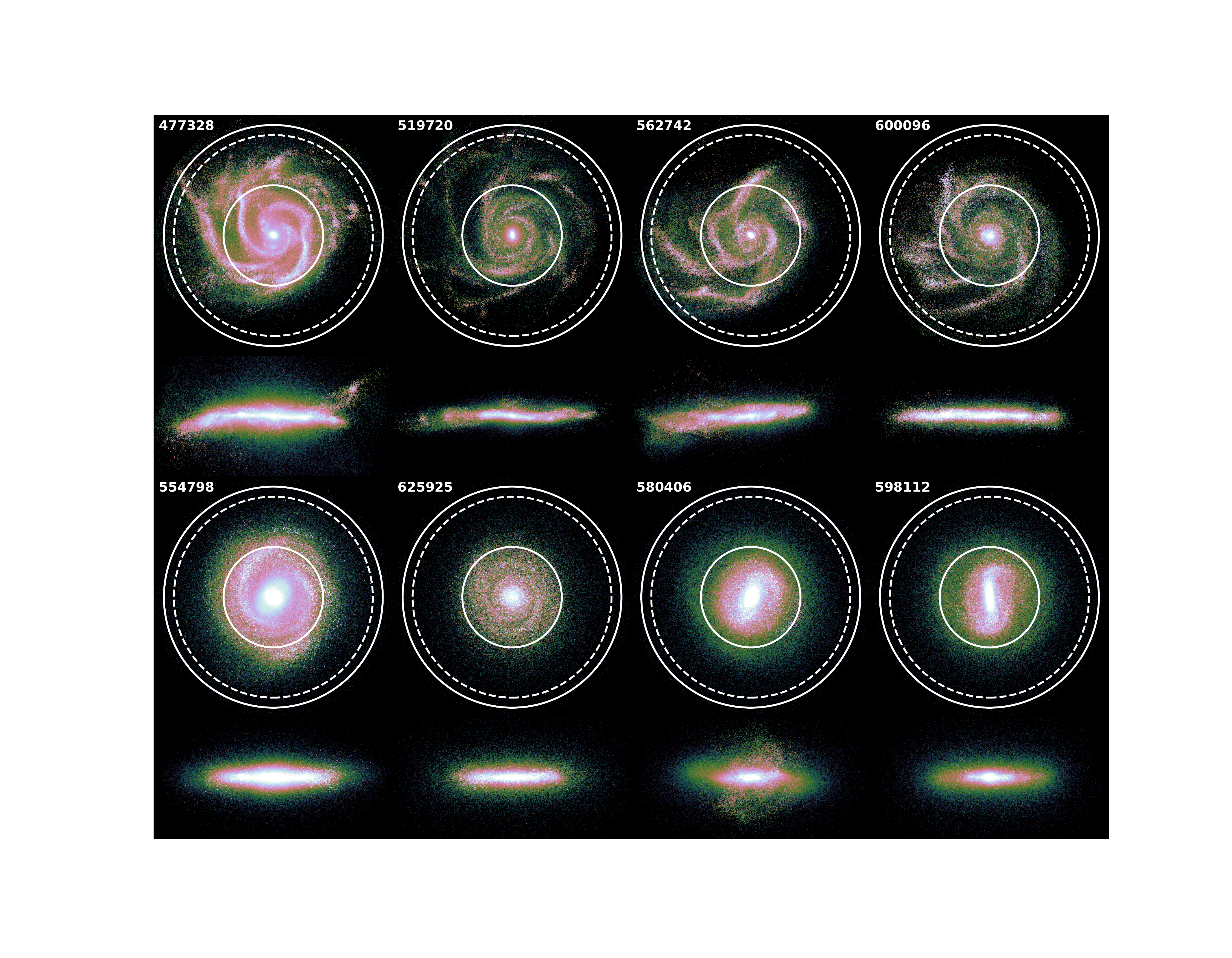}
\caption{Face-on and edge-on projected stellar density for eight galaxies from our TNG50 sample at $z=0$. The images at the top correspond to the most lopsided galaxies while those at the bottom to the most symmetrical ones. The dashed circle indicates \Rgal. The solid circles represent the lower and upper radial limit considered to compute \Am~, which are between 0.5\Rgal\ and 1.1\Rgal, respectively. More details in Section \ref{fourier}. }
\label{galaxies_example}
\end{center}
\end{figure*}

\section{Simulations}
\label{sec:simulation_data}

In this Section we introduce the numerical simulations considered in this work, which are taken from the  Illustris-The Next Generation project  \citep[IllustrisTNG hereafter][]{Pillepich2018b, Nelson2018, Nelson2019a, Marinacci2018, Springel2018, Naiman2018}. We also describe the criteria applied to select galaxies from the  corresponding large  cosmological boxes.

\subsection{The IllustrisTNG proyect}

The IllustrisTNG project is a set of gravo-magnetohydrodynamics cosmological simulation, run  with the moving-mesh code Arepo  \citep{Springel2010}.  It comprises
three large simulation volumes: TNG50, TNG100, and TNG300, enclosing volumes of $\sim 50^3$ cMpc, $100^3$ cMpc and $300^3$ cMpc, respectively. All these TNG runs follow the standard $\rm \Lambda$CDM model, with parameters based on the \cite{PlanckCollaborationXIII} results:  $\Omega_m$= 0.3089,  $\;\Omega_\Lambda$= 0.6911,$\;\Omega_b$= 0.0486, h= 0.6774,$\;\sigma_8$= 0.8159,$\;n_
s$= 0.9667, with Newtonian self-gravity solved in an expanding Universe. The IllustrisTNG\footnote{https://www.tng-project.org} is the successor of the Illustris project \citep{Vogelsberger2014Natur,Vogelsberger2014,Genel2014,Nelson2015A&C}, containing updated models for the physical processes that are relevant for galaxy formation and evolution \citep{Weinberger2017,Pillepich2018b}, 
 such as radiative cooling, stochastic star-formation in dense interstellar stellar medium, and an updated set of sub-grid physics models for stellar evolution, black hole growth,  stellar and AGN feedback.

In this work, we  focus on the model TNG50-1 \citep[][]{Pillepich2019MNRAStng50,Nelson2019MNRAStng50} and its DM only counterpart. TNG50-1 (hereafter TNG50) is the highest resolution run within the TNG project. Its high resolution allows us to better analyze the azimuthal distribution of stellar mass in the outskirts of Milky Way-like galaxies. In Table \ref{tablesimu} we list the main parameters of this simulation. 

The TNG50 database provides a catalogue of magnitudes in eight bands (SDSS g,r,i,z, Buser U,B,V and Palomar K) for each stellar particle. To estimate them, each stellar particle is assumed to represent a single stellar population of a given age and metallicity,  consistent with a Chabrier IMF \citep{Chabrier_2003}. Their energy spectral distributions (SEDs) are obtained from the \cite{Bruzual2003} populations synthesis models \citep[e.g.][]{Tissera1997}. We note that possible effects by dust obscuration have not being considered.

\subsection{Selection criteria}
\label{sec:crit}

In this work, we seek to characterize the properties and main physical mechanism  that give rise to disc galaxies displaying a non-axisymmetric global mass distribution of type $m=1$, better known as lopsided galaxies \citep{Jog2009}. 

We built our sample focusing on host late-type galaxies embedded in DM haloes with a \Mcrit\; between $10^{11.5}$ and $10^{12.5}~M_{\odot}$, where \Mcrit\;is defined as the total mass of the halo enclosed in a sphere whose mean density is 200 times the critical density of the Universe at $z=0$.  We considered only central galaxies, so we do not consider satellites within our sample. To properly quantify  lopsidedness in the galaxies outskirts, we selected well-resolved galaxies with more than $ 10000$ stellar particles, identified and assigned to each host by the  SUBFIND algorithm \citep{Springel2001}. Finally, we selected disc-dominated galaxies by requiring the Disc-to-Total mass ratio (D/T) to be greater than 0.5.
This last parameter was extracted from a catalogue provided by \cite{Genel2015}, and represents the fractional stellar mass within 10$\times R_{50}$\footnote{The stellar half-mass radius, $R_{50}$ is defined as the radius that encloses 50 percent of the total stellar mass of a subhalo.} with a circularity parameter $|\epsilon| >0.7$. That last parameter is defined as $\epsilon = J_z / J(E)$, where $J_z$ is the angular momentum component perpendicular to the disc plane of a stellar particle with orbital energy $E$, and  $J(E)$ is the (estimated) maximum possible angular momentum for the given E in a circular orbit \citep{Tissera2012}. That last selection about D/T place a strong limit on the mass contribution of the spheroidal components to the simulated galaxies.

After applying the selection criteria,  the final sample comprises 240 late-type galaxies at $z=0$. In Figure \ref{statesample_Mstar}, we show the total stellar mass distribution of the selected sample. The stellar mass distribution of our TNG50 sample ranges from $10^{9.5}$ M$_{\odot}$ to $10^{11.2}$ M$_{\odot}$. The mean stellar mass of our galaxy sample is $\sim 10^{10.3}$ M$_{\odot}$. In Section \ref{sec:driving}, we expand our sample to compare with previous results from the literature. Only for this purpose, we select central haloes with $M_{200}$ ranging from 10$^{11}$ to 10$^{13}$ $M_{\odot}$. 


\begin{table}

   \begin{center}
   \caption{ Main parameters of the TNG50 simulation:
   the comoving volume and the box side-length (1-2$^{\rm th}$ rows), the number of initial gas cells and dark matter particles (2-4$^{\rm th}$ rows), the mean baryonic and dark matter particle mass resolution (4-6$^{\rm th}$ rows), the minimum allowed adaptive gravitational softening length for gas cells (comoving Plummer equivalent) (7$^{\rm th}$ row) and  the redshift zero softening of the collisionless components (8$^{\rm th}$ row).}
   \label{tablesimu}
\begin{tabular}{lccc}
\hline
\textbf{Run Name}     & \multicolumn{1}{l}{\textbf{}} & \multicolumn{1}{l}{\textbf{TNG50}}  \\ \hline
Volume                & {[}$\rm cMpc^3${]}             & $51.7^3$                       \\ 
$\rm L_{box}$             & {[}cMpc/$h${]}                 & 35                                \\ 
$\rm N_{GAS}$             & -                             & $\rm 2160^3$                      \\ 
$\rm N_{DM}$              & -                             & $\rm 2160^3$                      \\ 
$\rm m_{baryon}$          & {[}$\rm M_{\odot}${]}          & $\rm 8.5 x 10^4$                  \\ 
$\rm m_{DM}$              & {[}$\rm M_{\odot}${]}          & $\rm 4.5 x 10^5$                  \\ 
$\rm \epsilon_{gas,min}$  & {[}pc{]}                      & 74                                \\ 
$\rm \epsilon_{DM}$       & {[}pc{]}                      & 288                                \\
\hline
\end{tabular}
\end{center}
\end{table}

\section{Methods}
\label{sec:methods}
\subsection{Characteristic scales}

To measure asymmetries in the mass and light distribution of the disc component of our simulated galaxy suite, it is important to define the different radial scales within which  the analysis will be performed. In our work, these characteristic scales are estimated by using the projected  stellar mass and light onto the rotational plane of the disc. First, we generate radial surface brightness (SB) profiles in the V photometric band. The SB profiles are created through the binning of the luminosity distribution of the stellar particles in radial annuli of 0.5 kpc of width. For better accuracy, we have smoothed the SB profile with a polynomial fit. This smoothed profile is used to define the position outermost edge of the disc, \Rgal, as the radius where the SB profile falls to a magnitude of 26.5 $\rm mag\;arcsec^{-2}$. \Rgal\ is also known as optical radius, and here it is used  as a proxy of the size of galaxies. The \Rgal~ in our TNG50 sample are within the range [9.5, 46.75] kpc with a median of 22.53 kpc. In Figure \ref{galaxies_example} we show examples of  four lopsided  and four symmetric discs galaxies in our sample  (top and bottom panel, respectively). In this figure, we also illustrate the sizes of the galaxies as measured by \Rgal~ (white dashed circles), illustrating how well this parameter traces the size of disc galaxies with different characteristics.

From now on,  we consider all star particles located within a sphere of radius \Rgal, to estimate the parameters in this subsection. We define the  stellar half-mass radius, \Rhalf~ as the position that enclosed the 50 percent of stellar mass, $\rm M^{\star}_{50}$ of the corresponding disc. Similarly, we define \Rt~ as the position that enclosed 90 percent of the disc stellar mass, $\rm M^{\star}_{90}$.  We find that \Rhalf~ varies  between 2.02 and 13.69 kpc with a median of 6.57 kpc, while \Rt\;  varies within 5.91 and 36.26 kpc, with a median of 15.81 kpc. These parameters allow the estimation of the stellar concentration defined as $C_\star=R_{\rm 90}^{\star}/R_{\rm 50}^{\star}$, and central stellar density, $\mu_\star=M^{\star}_{\rm 50}/ \pi {R^{\star}_{\rm 50}}^2$.

\begin{figure}
\begin{center}
\includegraphics[trim={0cm 0cm 1.5cm 1.5cm},clip,width=\linewidth]{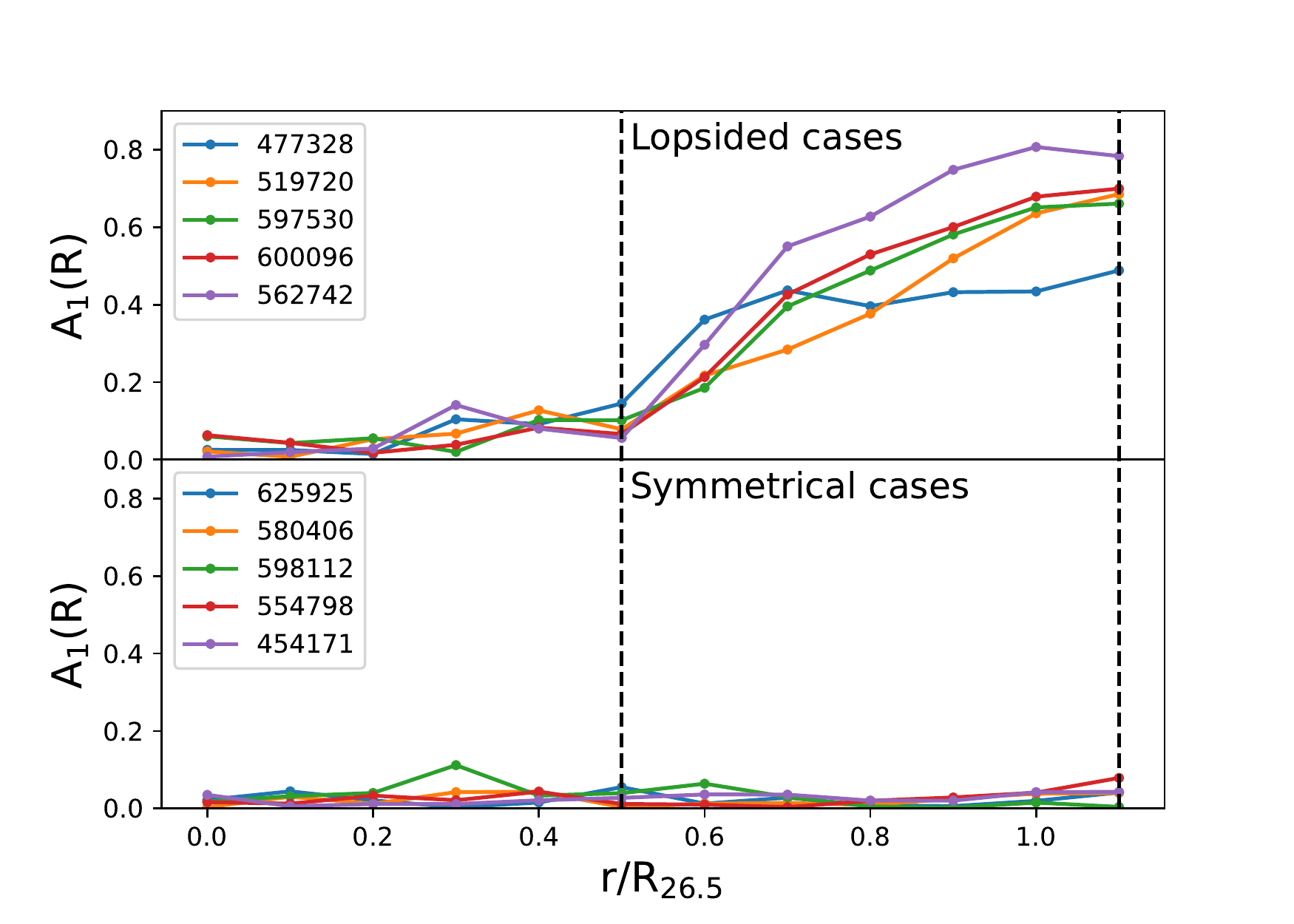}
\includegraphics[trim={0cm 0cm 1.5cm 1.5cm},clip,width=\linewidth]{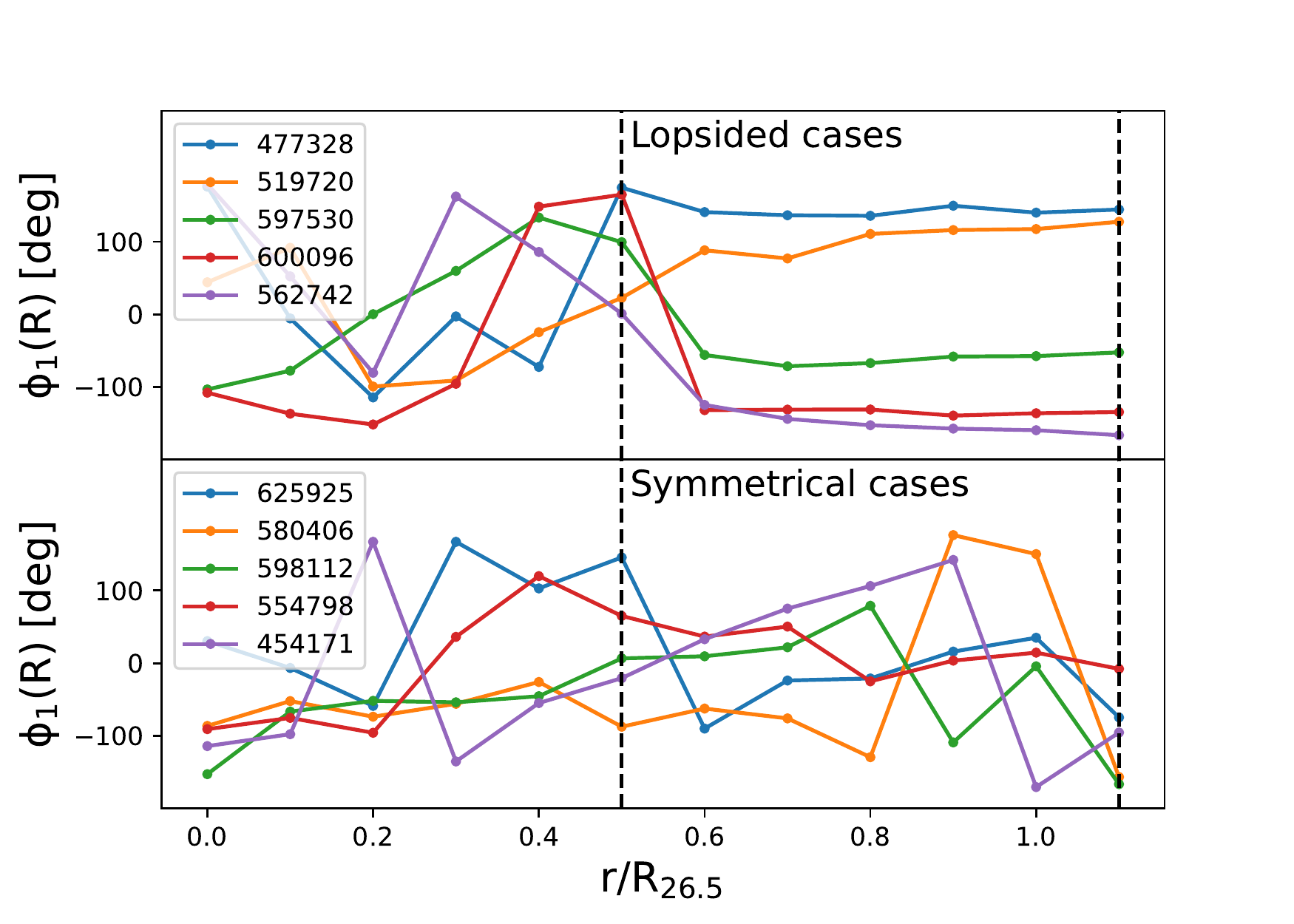}
\caption{{\it Top panel:} Radial distribution of the amplitude $m=1$ Fourier mode (\Am($R$)), for five galaxies lopsided   of \Am (top sub-panel) and five symmetric galaxies (bottom sub-panel). The black  dashed lines indicate the lower and upper radial limits considered to compute a global mass-weighted mean of $m=1$ Fourier amplitudes, \Am\ for each galaxy. Note that lopsided cases show an increase of \Am($R$) when increasing radius, and for symmetrical cases \Am($R$) approaches zero in the whole galaxy (see Section \ref{fourier} ). {\it Bottom panel:} Radial distribution of phase angle of the $m=1$ component, $\phi_1(R)$. Similar to the top panel, we show five lopsided and symmetrical disc examples. The lopsided galaxies show a nearly constant phase angle in their outer disc, in agreement with previous studies \citep{Li2011,Zaritsky2013}.}  \label{A1sample}
\end{center}
\end{figure}

\subsection{Quantification of $m=1$ asymmetries}
\label{fourier}
In order to quantify the asymmetry in the mass and light distributions of the  disc stellar component of our galaxies, we adopt the Fourier mode approach \citep{Rix1995,Zaritsky_1997,Eymeren2011,grand2016,Quillen2011}. In particular, we focus our analysis on lopsided perturbations, which can be characterized as a displacement of the center of stellar mass with respect to its center of density. Such asymmetric perturbations can be quantified through the amplitude of the $m=1$ Fourier mode.

Within a given thin radial annulus, $R_{\rm j}$, the complex coefficients of the $m$ Fourier mode  can be  estimated from a discrete distribution as

\begin{equation}
    \label{eqCoefF}
     C_{\rm m}(R_{\rm j})=\sum^{\rm N} _{\rm i} M^*_{\rm i}e^{-{\rm i} m\phi_{\rm i}}
\end{equation}

where $M^*_i$ and $\theta_i$ are the mass and azimuthal coordinate of the $i$-th stellar particle that belongs to the $j$-th radial annulus in a given galaxy. the angle $\phi_i$ is defined as $\phi_i = {\rm atan2}(y_i,x_i)$, where $x_i$ and $y_i$ are the cartesian coordinates of the i-th stellar particle for galaxies oriented in a face-on configuration\footnote{atan2() is a function of two parameters that returns the phase angle of the position of a i-th star particle in the respective quadrant, thus phase angles have values within the range $-\pi$ to $\pi$.}. 
Then we define the amplitude  of $m$-th Fourier mode as,
\begin{equation}
\label{eq3}
    B_{\rm m}(R_{\rm j})=\sqrt{a_{\rm m}(R_{\rm j})^2 + b_{\rm m}(R_{\rm j})^2}
\end{equation}
where $a_{\rm m}$ and $b_{\rm m}$ are the real an imaginary part of $C_{\rm m}$ (equation \ref{eqCoefF}). The amplitude $B_{\rm 1}(R_{\rm j})$ corresponds to the strength of the $m=1$ mode within a given $j$-th radial annulus. Finally, since each radial annulus has a different  total stellar mass, we express  $B_{\rm 1}(R_{\rm j})$ relative to the corresponding $m=0$ mode,

\begin{figure}
\begin{center}
\includegraphics[trim={0.5cm 0cm 1.5cm 1.5cm},clip,width=\linewidth]{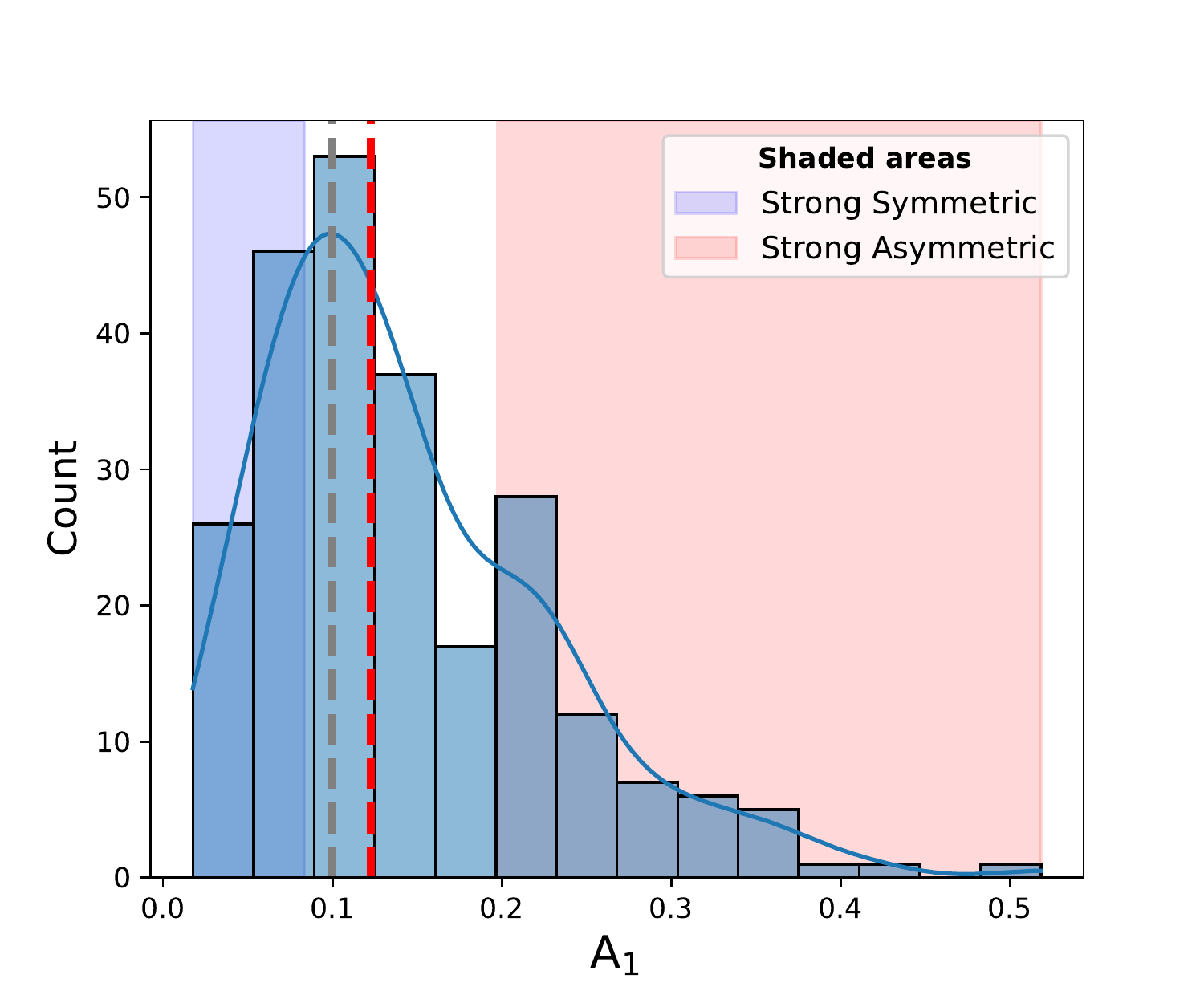}
\caption{Distribution of the sample of global (mean) \Am\ parameter computed for each simulated galaxy in our sample at $z=0$. The solid blue line shows the \Am~ distribution built using KDE method. The dashed red line is the median of \Am~ distribution, $\hat{A}_1 \approx 0.12$, which is used to differentiate between symmetric and asymmetric galaxies, the gray dashed line correspond to 0.1 threshold, typically used to defined lopsided galaxies, note that $\hat{A}_1>0.1$, this means that a little more than half of our sample has $\rm A_1>0.1$ values. The blue and red shaded areas indicate the first and fourth quartiles of the distribution, used to define the sub-sample of strong symmetric and asymmetric groups, respectively.}
\label{statesample_A1}
\end{center}
\end{figure}

\begin{equation}
\label{eqA1}
     A_{\rm 1}(R_{\rm j}) = \frac{B_{\rm 1}(R_{\rm j})}{B_{\rm 0}(R_{\rm j})},
\end{equation}
where $B_{\rm 0}$ is given by eqn. \ref{eq3} for $m=0$, and it is equal to the total mass in the given $j$-th radial annulus. Thus,  eqn. \ref{eqA1} corresponds to the mass-weighted amplitude of the $m=1$ Fourier mode as a function of radius.   

In Fig. \ref{A1sample} we show the radial $A_1$ profile (top panel) obtained from  five of our most lopsided  (top sub-panel) and five of our most symmetrical models (bottom sub-panel). We note that all galaxies, independently of whether they are lopsided or not, show very small $A_1$ values within  $R \sim 0.5$ \Rgal. However, for lopsided galaxies, $A_1$ starts to rapidly increase after this galactocentric distance. The radial distribution of \Am$(R)$ in our model is similar to that found in \cite{Rix1995}, who used near-IR observations from a sample of 18 galaxies  to characterise the properties of lopsided galaxies. \citet{Rudnick1998} and \citet{Bournaud2005} also found that the amplitude of the lopsided perturbations increases steadily  ($A_1 > 0.1)$ within the outer disc regions (radial range of $\approx 1.5$ to $2.5$ exponential disc scalelenghts).  \cite{Jog2000} suggested that the self-gravitational potential of the galaxy exerts a resistance to some external gravitational perturbation. However, the resilience exerted by self-gravity is more significant at smaller radii, and indeed the values of \Am(R) are low in the inner disc. For lopsided galaxies, the gravitational pull by self-gravity is weaker at larger radii, so \Am(R) grows. Otherwise, the symmetric cases could be gravitationally more cohesive, and consequently the radial distribution of \Am(R) keeps lower values in the whole disc. In Section \ref{subtidal}, we explore this in detail. 

In the bottom panel of Fig. \ref{A1sample}, we show the radial phase angle of the $m=1$ component, $\phi_{\rm 1}(R)$. Note the nearly constant value of $\phi_1$(R) in the outer disc for lopsided examples, region where the corresponding  asymmetry becomes significant. This feature is typical in lopsided galaxies \citep{Zaritsky_1997,Eymeren2011,Ghosh2022}. The radial variation of $\phi_{\rm 1}(R)$ is a useful tool for understanding the nature of the lopsidedness and how long it takes to wind around the galaxy \citep{Baldwin1980}. Previous  results \citep{Saha2007, Ghosh2022} suggest that, in lopsided galaxies, the outer galaxy region  does not wind up as quickly as their inner region, suggesting a weak self-gravity in these galaxies.

Since the outer region of galactic discs is more prone to developing lopsidedness, we estimate, for each galaxy, a unique global mass-weighted mean of the $m=1$ Fourier mode, hereafter \Am. This allow us to compare the level of lopsidedness among galaxies in our sample. The global \Am\; is computed by taking the mean of the \Am($R$) in outer galaxy regions. We consider eight (8) radial annular region, of width 0.075\Rgal, located within the interval 0.5\Rgal~ to 1.1\Rgal. This region is highlighted by the dashed lines in Fig. \ref{A1sample} and the solid circles shown in Fig. \ref{galaxies_example}).

\begin{figure}
\begin{center}
\includegraphics[trim={0cm 0cm 1cm 1cm},clip,width=\linewidth]{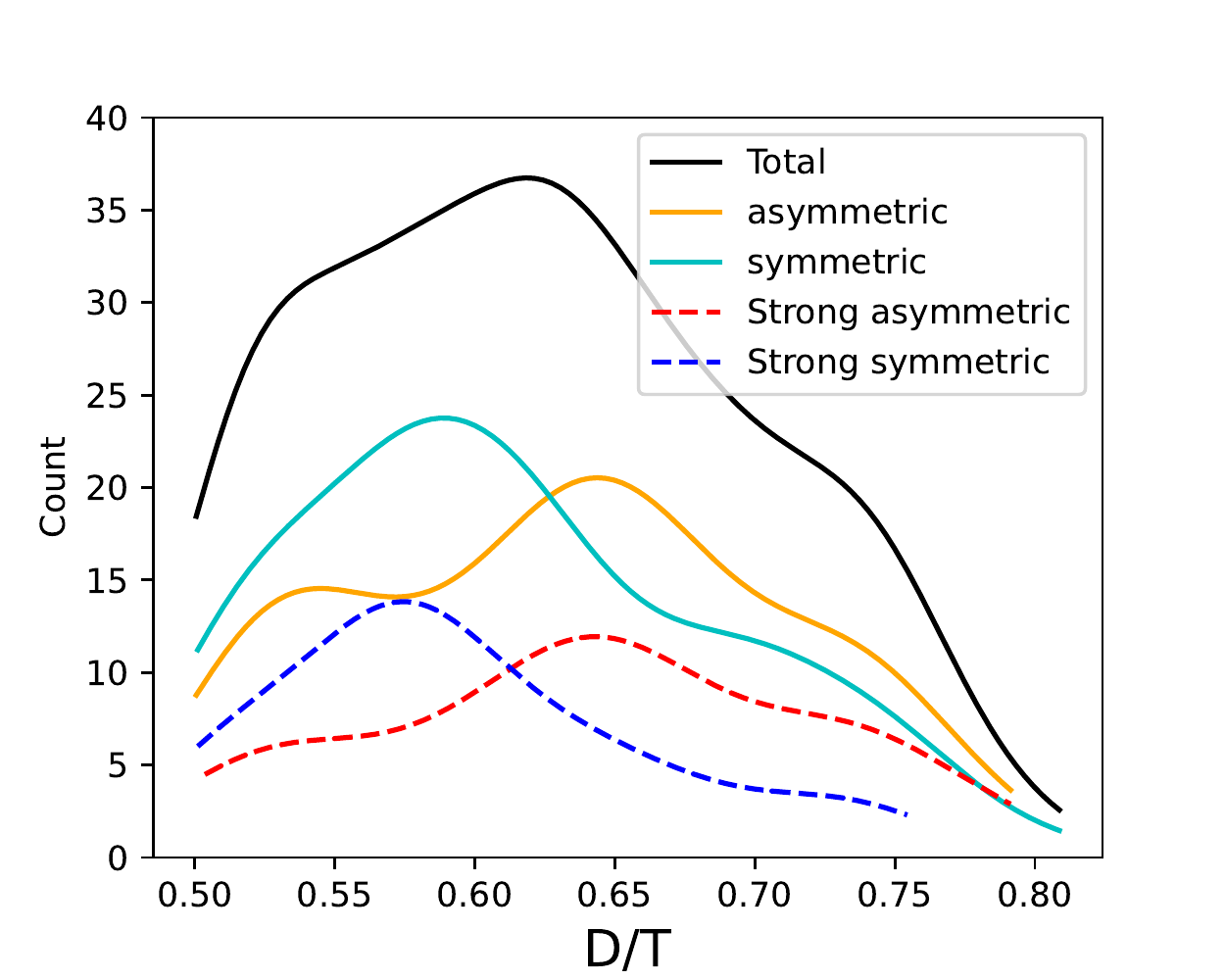}
\caption{Distribution of Disc-to-Total mass ratio, D/T, for the total sample (black line), only asymmetrical galaxies (orange line) and only symmetrical galaxies (cyan). The red and blue dashed lines correspond  to strong asymmetric and symmetric subsamples, respectively. The distribution were obtained using a KDE method. Asymmetrical galaxies tend to be more disc-dominated than their symmetrical counterpart. The medians for each group are 0.61 and 0.64, respectively. A similar trend is observed for the strong sub-samples.}\label{DTdistribution}
\end{center}
\end{figure}

\subsection{Estimating the asymmetries in DM haloes and stellar component}
\label{MeasureDMCM}

In this work we are interested in characterizing the origin and evolution of lopsided perturbations. A possible mechanism triggering such perturbation is the response of the galactic disc to a distorted DM halo. As discussed in Sec.\ref{sec:intro}, these halo distortions arise as a  result of interactions between the host DM particles and an external agent  \citep{Jog1997,Jog1999,Gomez2016,gao,Laporte2018}. To quantify such distortions in the DM halo of our numerical models, we focus on offsets of the halo center of mass with respect to is density cusp. Typically, the dipolar response of the DM halo density field is the strongest. Thus, it can be used to identify perturbed DM distributions.

Here we follow the analysis performed by \cite{gao}. First, we identify the DM halo density cusp, $r_{\rm cusp}$, based on the position of the most bound particle of the central halo, given by SUBFIND. We then computed the DM halo center of mass,  $r_{\rm DM}$, considering all DM particles located within the inner three and five times \Rgal\; We note that, as shown  by \cite{Gomez2016}, perturbations in the DM halo at further galactocentric distances  are not efficient at disturbing the embedded galactic disc. Nonetheless, to compare with \cite{gao} we also compute $r_{\rm DM}$ considering all DM particles assigned to the main host halo by SUBFIND ($R_{200}$). Finally we compute the offset of $r_{\rm DM}$ with respect to $r_{\rm cusp}$ as   
\begin{equation}
\label{eq:delta_rdm}
      \Delta r^{i}_{\rm DM} =  |r_{\rm cusp} -r^{i}_{\rm DM}|,
\end{equation}
where the supra index $i = 3R_{26.5}, 5 R_{26.5}$ and $R_{\rm 200}$ indicates the spatial region within which  $r_{\rm DM}$ is calculated.

\label{PhysicalSec}

\section{Results}
\label{sec:results}

\subsection{General disc morphological properties}
\label{Properties}

In this Section we analyze the main morphological characteristics of the 240 stellar discs, selected according to the criteria defined in Section \ref{sec:simulation_data}.
In Fig. ~\ref{statesample_A1} we show the distribution of the global \Am~ parameter, which correspond to the average $A_1(R)$ values computed within the radial range $[0.5-1.1R_{26.5}]$ (see Sec. \ref{fourier}). We note that the distribution is similar to the one reported by \citetalias{Reichard2008} (see their figure 10), obtained using a sample 25155 galaxies from the SDSS. It is worth noting that, even though the stellar mass range of our sample (Fig. \ref{statesample_Mstar}) is similar to that in \citetalias{Reichard2008} ($10^8 - 10^{11} M_\odot$, see figure 8 in \citetalias{Reichard2008}), the latter includes a population of early type galaxies, which are missing from our sample. Nonetheless, the \citetalias{Reichard2008} sample is dominated by late-type objects, allowing us to compare  our results with the data. The characteristic galaxy \Am\;values in \citetalias{Reichard2008} were obtained by averaging over the radial range between $R_{50}$ and $R_{90}$. The outer radius limit is imposed due to limitation with the observational data (see Sec. 2.2 of \citetalias{Reichard2008} for more details). We have computed our distribution considering  smaller outer limits, finding no significant variation in our results. Similar results were obtained by previous works such as \citet{Rix1995, Bournaud2005}.

The red dashed line in Fig. \ref{statesample_A1} indicates the median of the $A_1$ distribution, which takes a values of $\hat{A}_1 \approx 0.12$. This $\hat{A}_1$ is used from now on to differentiate  galaxies between symmetric ($A_{1} < \hat{A}_1$) and asymmetric or lopsided cases ($A_{1} > \hat{A}_1$). We note that this value is only slightly larger than the 0.1 threshold, typically used to define lopsided discs \citep{Bournaud2005,Zaritsky_1997,Jog2009}. We further subdivide our sample into strongly symmetric and asymmetric cases by selecting galaxies located in the first and fourth quartiles of the $A_{1}$ distribution, respectively. The strong cases are highlighted in Fig. \ref{statesample_A1} with shaded areas.

 We now explore whether there are correlations between the  D/T (see Sec. \ref{sec:crit}) of our simulated galaxies and the symmetry of their azimuthal mass distribution. We recall that the parameter D/T allows the quantification of  the disc mass contribution to the galaxy's total stellar mass. The black solid line in Fig. \ref{DTdistribution} shows the D/T density distribution obtain using the Kernel Density Estimation (KDE)\footnote{We implement KDE using the {\sc gaussian\_kde} function from the {\sc scipy} library. More details can be found at https://scipy.org}  of the D/T values obtained from our full sample. Note that our selection criteria imposes a lower D/T limit of 0.50. The distribution has a median value of  $\approx 0.62$, indicating  a  significant presence of strongly disc-dominated  galaxies in our sample. Interestingly, asymmetric galaxies tend to be more disc dominated than their symmetric counterparts, with medians of  0.64 and 0.61, respectively.  The blue and red dashed lines show the same distribution, now for the strong-asymmetric and strong-symmetric samples. The difference in the median D/T values are slightly more pronounced than in the previous subsamples, with values of 0.64 and 0.58 for the asymmetric and symmetric subsamples, respectively. This suggests that the presence of more significant central pressure supported component could be playing a role on limiting the strength of lopsided perturbations. This is further explored in Section \ref{subtidal}.

\begin{figure}
\centering
\includegraphics[trim={0cm 0cm 1.5cm 1.5cm},clip,width=0.9\linewidth]{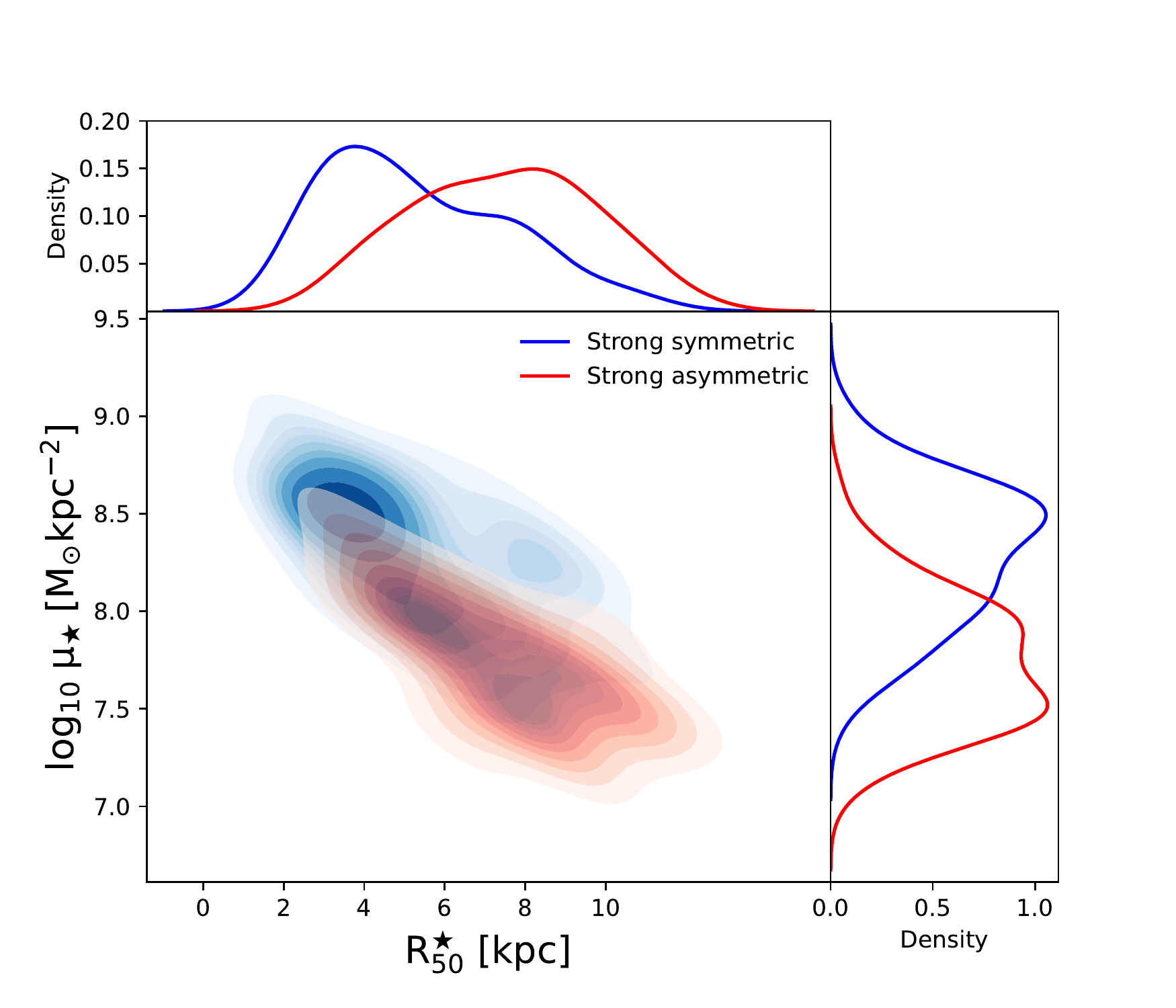}\\
\includegraphics[trim={0cm 0cm 1.5cm 1.5cm},clip,width=0.9\linewidth]{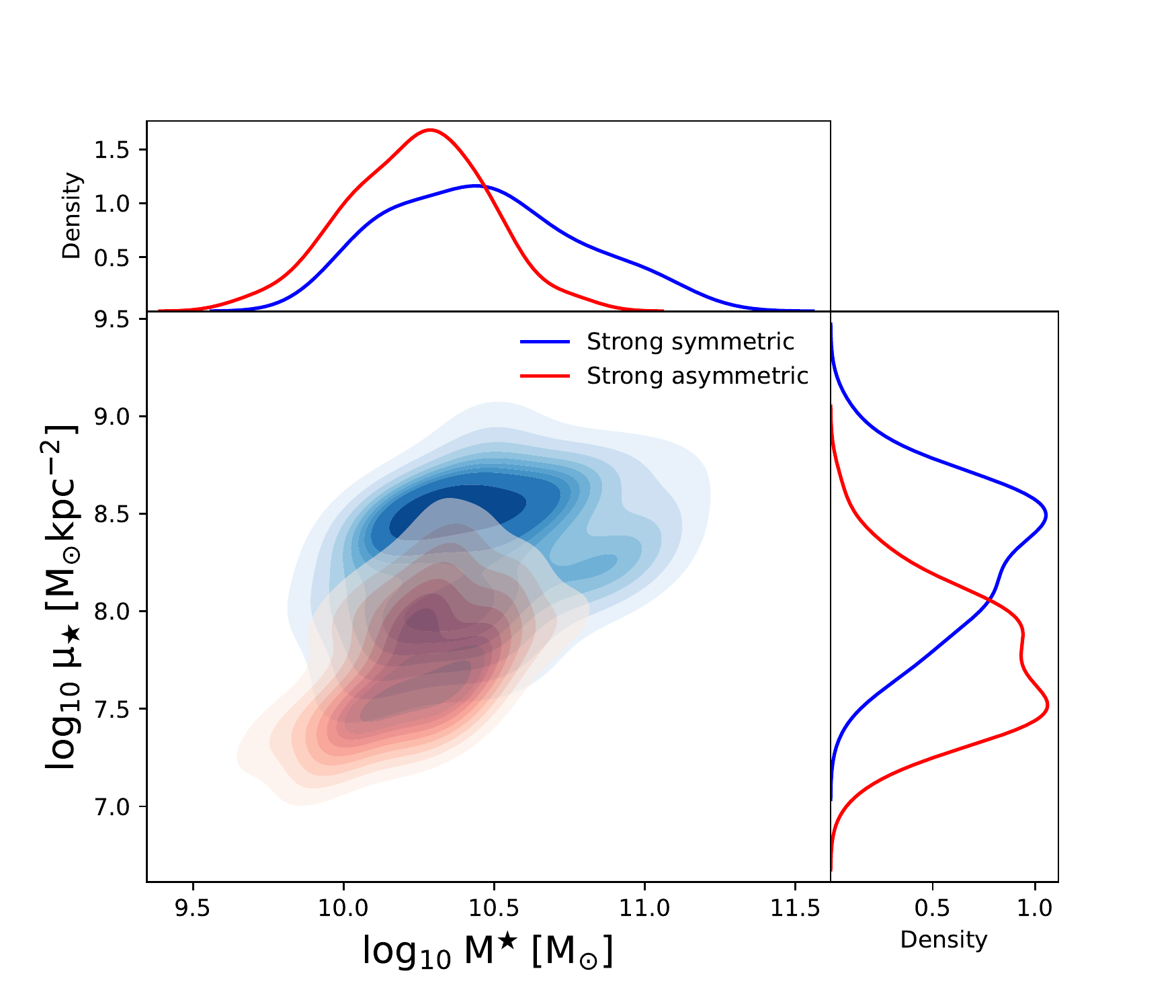}\\
\includegraphics[trim={0cm 0cm 1.5cm 1.5cm},clip,width=0.9\linewidth]{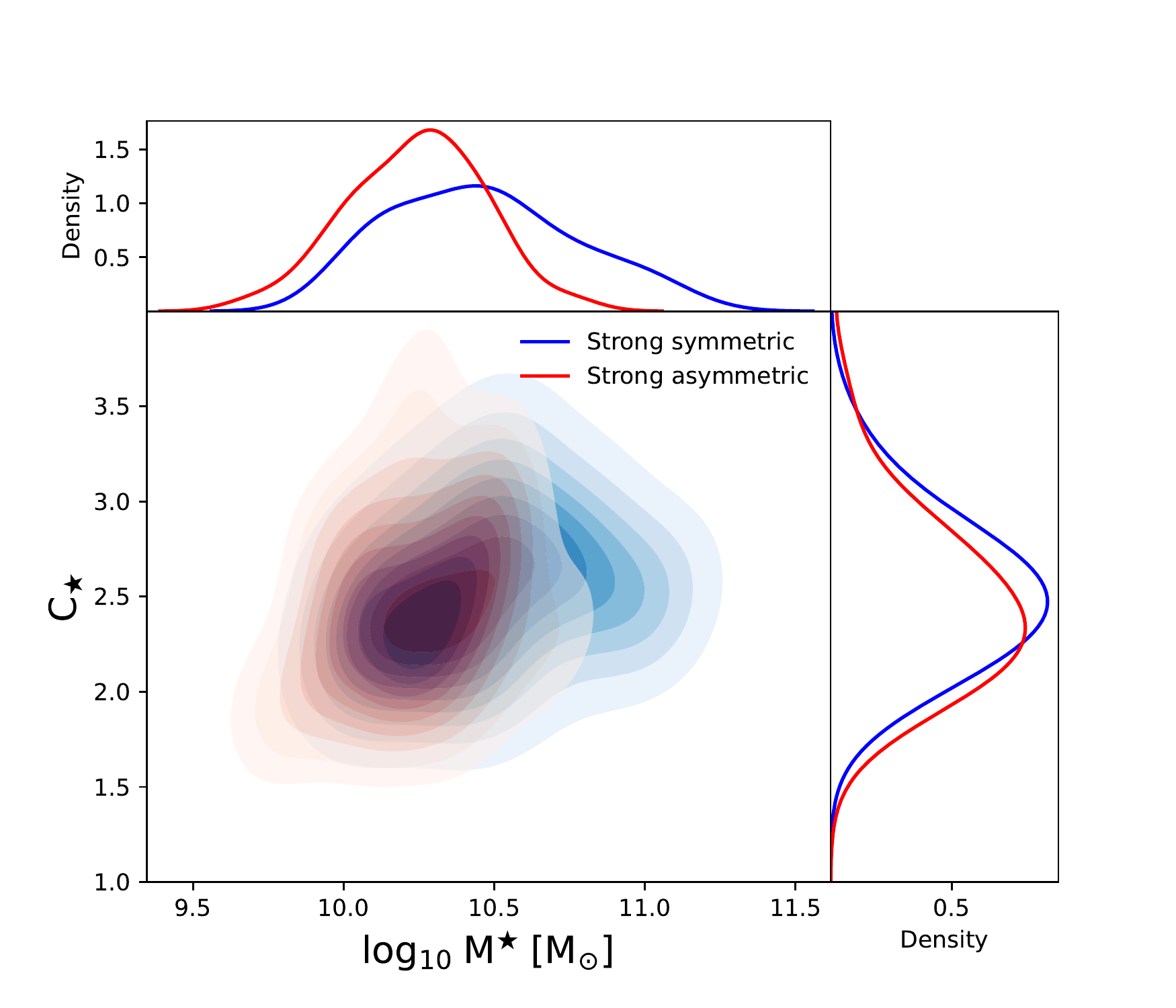}
\caption{ 
{\it Top panel:} Central stellar density, \mus, as a function of  stellar half-mass radius, \Rhalf, for the strong symmetric and asymmetric sub-samples that are defined in Fig. \ref{statesample_A1}.
{\it Middle panel:} \mus as a  function by the total stellar mass, \Mt.
{\it Bottom panel:}  Stellar mass concentration, $C_*$,  as a function \Mt.  These panels were built using KDE bivariate distribution for the central sub-panel, and simple KDE distribution for edge sub-panels. The strong sub-sample was separated between strong symmetric (blue region) and strong asymmetric (red region). Note that strong asymmetric galaxies tend to have their central regions more extended and slightly less massive than symmetric galaxies. Furthermore, A clear difference between both sub-samples are in their central stellar density, where in asymmetric galaxies tend to be lower ones.
}
\label{mainresult}
\end{figure}

\subsection{Structural properties of lopsided galaxies}
\label{comparizonasymm}

As discussed  in the previous Section, our sample of disc-dominated  galaxies show   different degrees of lopsided asymmetry, as quantified by the value of their A$_{1}$ parameter. In this Section, \Rhalf, stellar mass concentration, $C_{*}$, and central stellar surface density, \mus, present significant differences between the asymmetric and symmetric sub-samples.

To highlight the main differences between the asymmetric discs and their symmetrical counterparts, we focus on the strongly symmetric and strongly asymmetric galaxy samples defined in Fig. \ref{statesample_A1}. The top left panel of Fig. \ref{mainresult} shows the distribution of  $\rm R_{50}^{\star}$ and \mus. The distributions are represented with a two dimensional kernel density estimation (KDE). The top  and the right sub panels show the marginalized 1D distribution for $\rm R_{50}^{\star}$  and \mus, respectively . Interestingly, both strong types show different distributions in this plane. We find  that asymmetric galaxies tend to have larger \Rhalf~ than their symmetric counterparts. In addition, they tend to show lower values of \mus~ at given \Rhalf. These differences are highlighted on the 1D KDE, with median values of \Rhalf~ and \mus~ for the symmetric and asymmetric sub samples of $(4.62,7.22)$ kpc and $(10^{8.39},10^{7.75})\;\rm M_\odot \;kpc^{-2}$, respectively. 

Following \citetalias{Reichard2008}, on the middle and bottom panels we show  2D KDE of our simulated galaxy sample in \mus~  versus total stellar mass, $M^{*}$, and in $M^{*}$ versus stellar concentration, $C_{*}$, respectively. \citetalias{Reichard2008} shows that, among these structural parameters, the strongest correlation with $A_1$ is obtained for \mus. Indeed, our results are in good agreement with these observations. Note that the most pronounced difference between the distributions of these strong types is obtained for \mus. The marginalized $C_{*}$ distribution (bottom panel) shows that both type of galaxies present nearly  indistinguishable distribution of stellar concentration. Interestingly, within  the stellar mass range considered in this work, we find that symmetric galaxies tend to be slightly more massive than lopsided galaxies. In addition, the lopsided sub-sample shows a narrower distribution in $M^{*}$. To quantify these results, we estimate Pearson correlations coefficient between the previously defined parameters. In Table \ref{tabcorr}, we show both the correlation coefficient obtained using only the strong samples and also using all lopsided and symmetrical galaxies (fourth column). As previously indicated, the strongest (anti)correlation is obtained between \Am~ and $\rm \mu_\star$. This significant anti correlation is obtained for  both the strong  and the complete samples.

 \begin{table}
   \begin{center}
   \caption{{\bf Correlations coefficients for relations between the analysed parameters discussed in in Section \ref{PhysicalSec} and \ref{subtidal}.}
   }
 \label{tabcorr}
 \begin{tabular}{lrrr}\hline
 P1  & P2 & \multicolumn{2}{c}{Correlation Coefficient}\\ 
     &    &  Strong sample  &               All sample \\

 \hline
$\rm log_{10}$ \Am     & $\rm log_{10}\;\mu_\star$   &  -0.68 &  -0.54\\
$\rm log_{10}$ \Am     & \Rhalf            &  0.53 &  0.37\\
$\rm log_{10}$ \Am     & $\rm C_\star$     &  -0.03 &  -0.01\\
\hline\hline
$\rm log_{10}$ \Am     & $\rm log_{10}\;M_{50}/R^3_{26.5}$    &  -0.72 &  -0.60\\
$\rm log_{10}\;\mu_\star$     & $\rm log_{10}\;M_{50}/R^3_{26.5}$   &  0.60 &  0.52\\
 \hline
 \end{tabular}
  \end{center}
 \vspace{1mm}
 \end{table}

\subsection{The role of the central mass distribution}
\label{subtidal}

As discussed in the previous Section, our sample of galactic models shows a significant correlation between \Am~ and the central stellar density, \mus. Galaxies with lower \mus~  typically show higher values of \Am. Additionally we also find that lopsided galaxies tend to show larger values of \Rhalf. This suggest that galaxies with lower density and more extended central regions could be more prone to develop lopsided perturbations.  

\begin{figure}
\begin{center}
\includegraphics[trim={0.5cm 0cm 1.5cm 1.5cm},clip,width=\linewidth]{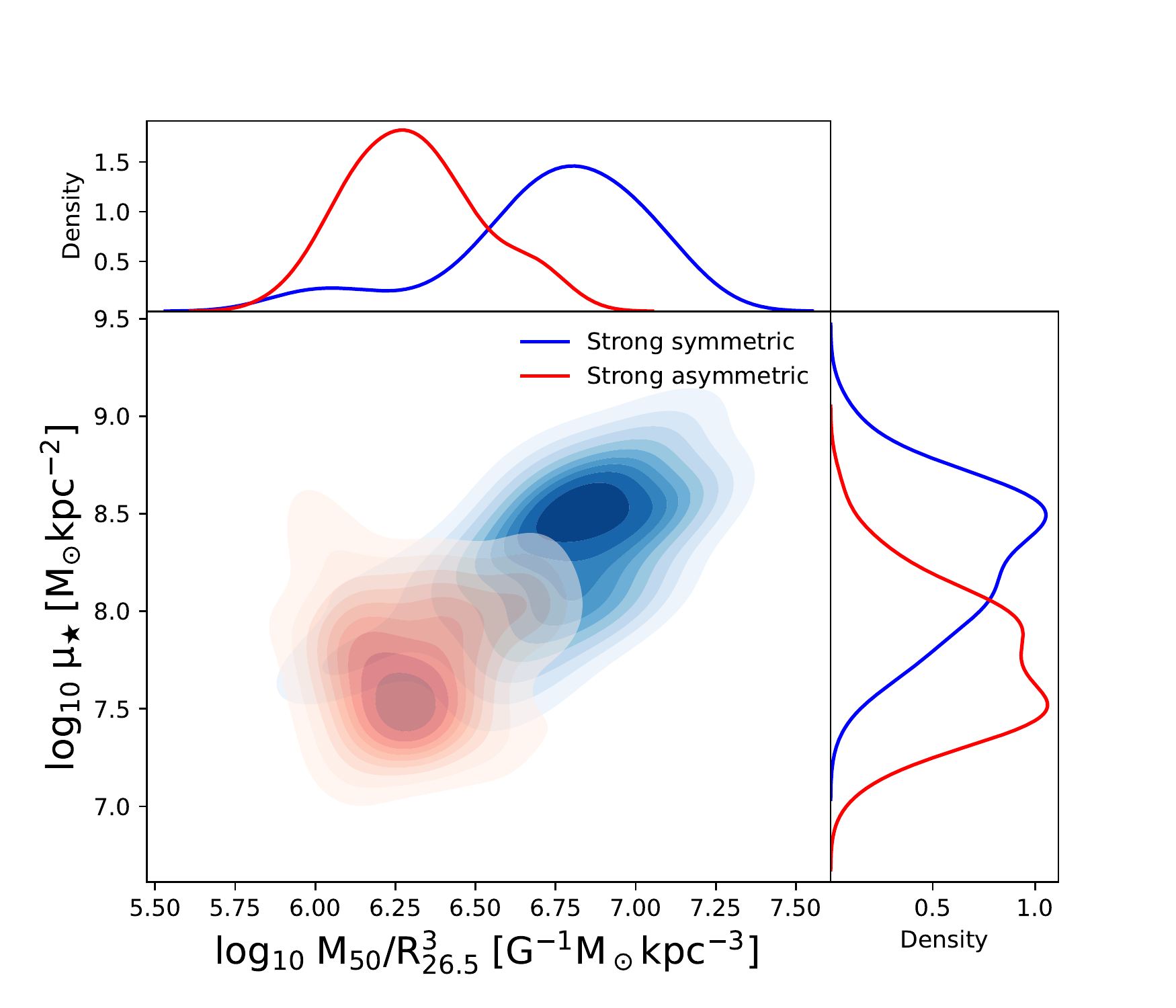}
\caption{ Distribution of the proxi of the tidal force exerted by the inner disc on the its outskirts, $\rm M_{50}/R^3_{26.5}$, versus central stellar density, \mus~,  for all galaxies in the strong symmetric and asymmetric samples. The top and right panels show the corresponding one dimensional distribution. All distributions where obtained using a KDE method. A clear correlation  between \mus and $\rm M_{50}/R^3_{26.5}$ is seen. Strong asymmetric galaxies tend to have a lower values of $\rm M_{50}/R^3_{26.5}$, suggesting that weakly gravitationally cohesive galaxies are susceptible to  lopsided distortions in their stellar distribution. }

\label{tidalforce}
\end{center}
\end{figure}

Indeed, discs with denser inner regions are likely to be more gravitationally cohesive. To explore whether the disc self-gravity plays a significant role in the onset and amplitude of lopsided perturbations we show, in Fig. \ref{tidalforce}, the distribution of  $\rm M_{50}/R^3_{26.5}$ versus \mus~  for all galaxies in the strong symmetric and asymmetric samples. Here, $\rm M_{50}$ represents the total mass of all particles enclosed in a sphere of radius \Rhalf, and includes  contributions from the stellar, the gas and the dark matter components. We note that the quantity $\rm M_{50}/R^3_{26.5}$ represents a proxy of the tidal force exerted by the inner galaxy region ($R < R^{\star}_{\rm 50}$) on material located a distances equal to the disc optical radius, $R_{26.5}$. As before, we focus on the strong symmetric and asymmetric types. From this figure we observe that these two subgroup represent very distinct populations in  $M_{50}/R^3_{26.5}$ versus \mus~ space. Present-day asymmetric galaxies exert a much lower tidal field on their outer disc regions, where lopsided perturbations show the strongest amplitudes. This is clearly shown on the 1D KDE histogram displayed in the top panel. Indeed, the (anti)correlation between \Am\ and  $M_{50}/R^3_{26.5}$ is the strongest among the structural parameters explored in this work. This is quantified in  Table \ref{tabcorr}, which also highlights that this anti-correlation is even greater than the one found between \Am~ and $\rm \mu_\star$. We emphasize that the anti-correlation is not limited to the strong subtypes, and that it remains large even if we consider the all galaxies in the sample, as can be seen from the rightmost column of Table \ref{tabcorr}.

 Several previous studies have explored different scenarios for the origin of lopsided modes based on environmental interactions, such as fly-bys, minor and major mergers, perturbed underlying dark matter density field, and misaligned accretion of cold gas, among others \citep{Weinberg1994,Jog1997,Jog1999,Kornreich2002,Walker1996, Zaritsky_1997,Bournaud2005,Levine1998,Noordermeer2001,Gomez2016,GaravitoCamargo2019}. Our results instead hint toward a population of galaxies susceptible to develop lopsidedness, and not to a particular external perturbation source. In other words, galaxies with weakly cohesive inner regions could develop a lopsided mode when faced with any sort of external perturbation. Indeed, as we show later in Section \ref{sec:driving}, the strong present-day connection between the strength of the lopsided modes and of the inner tidal force field is independent of the past interaction history of our simulated galaxies with their environment.

\subsection{Evolution of lopsided galaxies}

Around the 30 percent of late-type galaxies in the nearby Universe show lopsided perturbations \citep{Rix1995, Jog2009}. This could indicate that either lopsided perturbations are long lived, or that a significant fraction of galaxies are prone to develop such perturbations even in absence of significant external interactions, as suggested in the previous Section. Here we explore the time evolution of the main structural parameters that differentiate symmetric and lopsided galaxies, as well as the time evolution of the amplitude of the lopsided mode.

\subsubsection{Time evolution of structural parameters}
\label{sec:radiusintime}

\begin{figure}
\begin{center}
\includegraphics[trim={0cm 0cm 0cm 0cm},clip,width=0.95\linewidth]{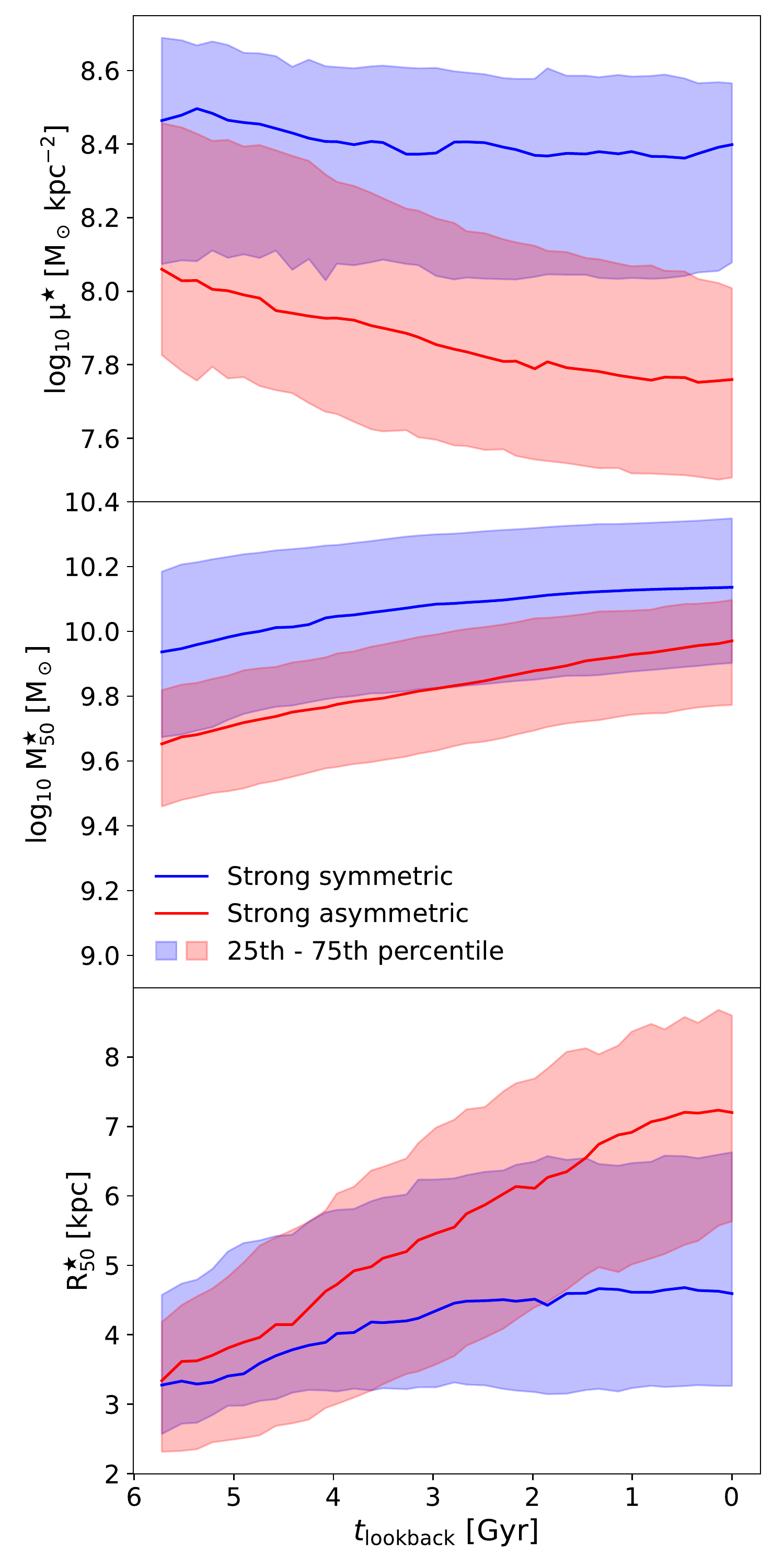}
\caption{Time evolution of the central stellar density, the stellar half-mass, and stellar half-mass radius (from top to bottom). Quantities are shown as a function of lookback time. The
solid blue and red lines show the median of the distributions obtained from the strong symmetric  and  asymmetric sumbsample, respectively. The shaded areas highlight indicate the 25th and 75th percentiles of the corresponding distributions. The central stellar density evolution of the strong asymmetric sub-sample tend to have a sharp decrease in time compared to strong symmetric sub-sample. This is a consequence of the rapid growth of \Rhalf\ over time. }\label{mur50m50_time}
\end{center}
\end{figure}

As previously discussed, lopsided galaxies tend to show lower values of \mus~ as well as larger \Rhalf~ at the present-day. In Fig.~\ref{mur50m50_time}  we explore how these  structural parameters evolved over time. As before we focus on the strong types to better highlight the differences between perturbed and unperturbed galaxies. The top panel shows the time evolution of  the central stellar surface density, \mus~, over the last 6 Gyr. The blue and red solid lines depict the median \mus~ obtained after stacking the strong asymmetric and symmetric subsamples, respectively. The shaded areas are determined by the 25th and 75th percentiles of both distributions. It is interesting to note that, at the present-day, the difference in \mus~ is very significant, and that this difference increased over the last 6 Gyr. In particular, we notice a significant decay of \mus~ over time for the strong lopsided cases, while for the symmetric counterparts \mus~ remain nearly constant. To understand what is behind this decay, we show in the middle panel of Fig. \ref{mur50m50_time}  the time evolution of the stellar mass enclosed within the stellar half-mass radius, \Mhalf. We notice that, on average, lopsided galaxies tend to enclose less stellar mass within \Rhalf. However the difference in \Mhalf~ between symmetric and asymmetric galaxies remains nearly constant over the last 6 Gyr. This is in contrast for the time evolution of \Rhalf, shown in the bottom panel. Note that 6 Gyr ago, both subsamples had, on average, very similar values of \Rhalf. However, lopsided galaxies experienced a significant growth of \Rhalf while, for symmetric galaxies, it remained nearly constant, specially over the last 3 Gyr. 

\begin{figure}
\begin{center}
\includegraphics[trim={0cm 0cm 1cm 1cm},clip,width=1.025\linewidth]{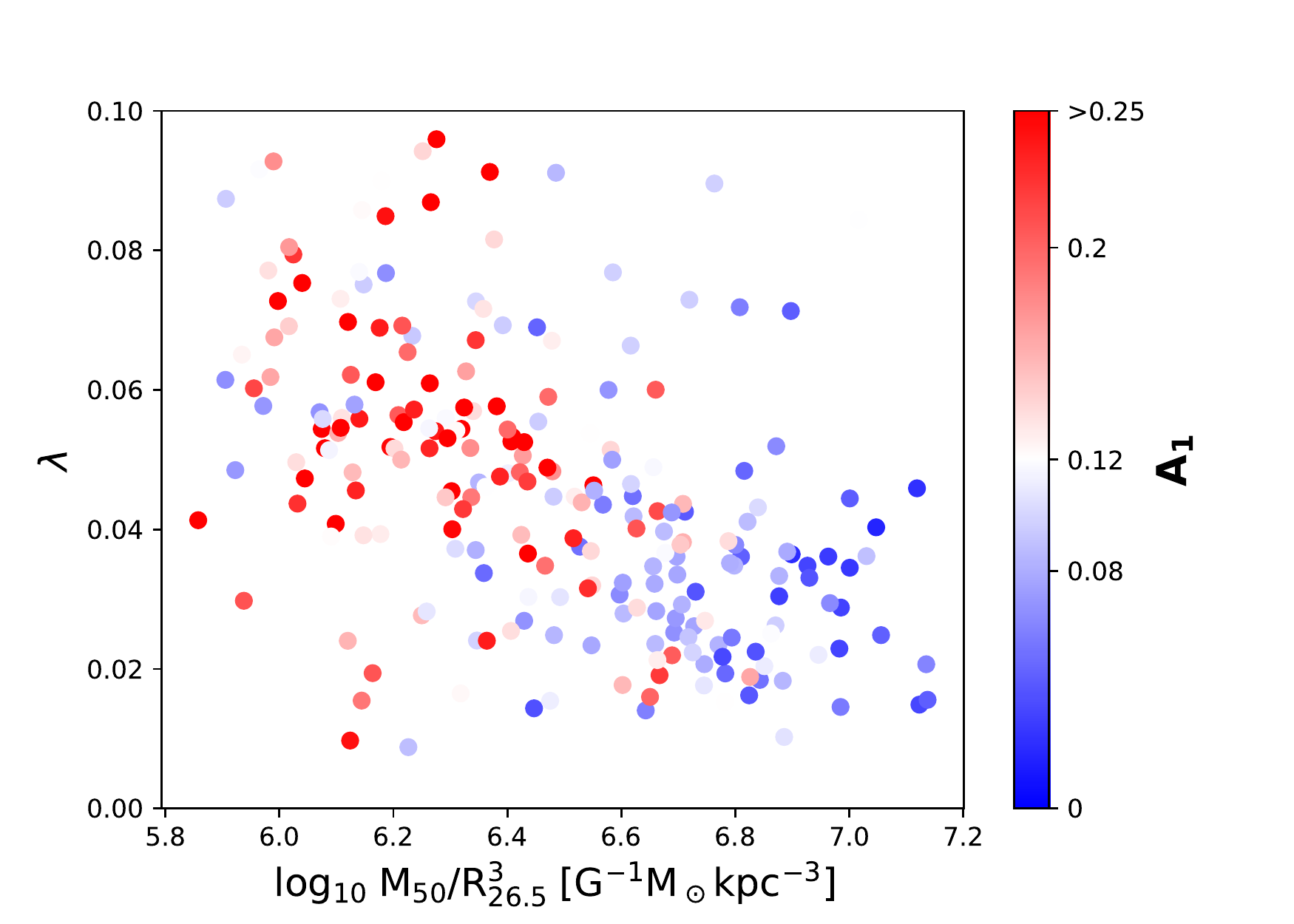}
\caption{ Distribution of halo spin parameter, $\lambda$, versus proxi of the tidal force exerted by the inner disc on the its outskirts, $\rm M_{50}/R^3_{26.5}$. The dots are colored according to the present-day value of \Am. Which the palette of colors was centered around $\hat{A}_1 \sim 0.12$, while 0.08 and 0.2 values correspond to the $25^{\rm th}$ and  $75^{\rm th}$ percentiles of \Am\; distribution, used to define strong sub-sample. Note that asymmetrical galaxies tend to have higher halo spin than their counterpart symmetrical.  }
\label{fig:lambda}
\end{center}
\end{figure}

The previous analysis shows that what drives the decay of \mus~ for lopsided galaxies is mainly the growth  of the stellar disc size. Using the Auriga simulations, \citet{Grand2017} investigated the mechanisms that set present-day disc sizes, and found that they are mainly related to the angular momentum of halo material. In their models, the largest discs are produced by quiescent mergers that inspiral into the galaxy and deposit high-angular momentum material into the pre-existing disc. This process simultaneously increases the spin of dark matter and gas in the halo. On the other hand early violent mergers and strong AGN feedback strongly limits the final  disc size by destroying pre-existing discs and by suppressing gas accretion on to the outer disc, respectively. Interestingly, they find that the most important factor that leads to compact discs, however, is simply a low angular momentum for the halo. To explore whether the halo spin, $\lambda$ \citep[see eq. 12][]{Grand2017} plays a role on the development of lopsided galaxies by partially setting the size of the disc and thus their radial mass distribution, in Fig. \ref{fig:lambda} we show the distribution of  $\lambda$ versus $M_{50}/R^3_{26.5}$ for galaxies in our sample. The color coding indicates the strength of the $A_1$ mode. Interestingly, we find that galaxies with high $\lambda$ typically show smaller values of $M_{50}/R^3_{26.5}$ and high values of $A_1$. On the other hand, galaxies with low $\lambda$ values are dominated by strongly self gravitating discs and, thus, low $A_1$ values. Interestingly, using the EAGLE and Fenix simulations \cite{Cataldi2021}  reported that haloes with less concentration tend to host extended galaxies. These results highlight an interesting morphology--halo connection for late type galaxies.

\begin{figure}
\begin{center}
\includegraphics[trim={0cm 0cm 0cm 0cm},clip,width=\linewidth]{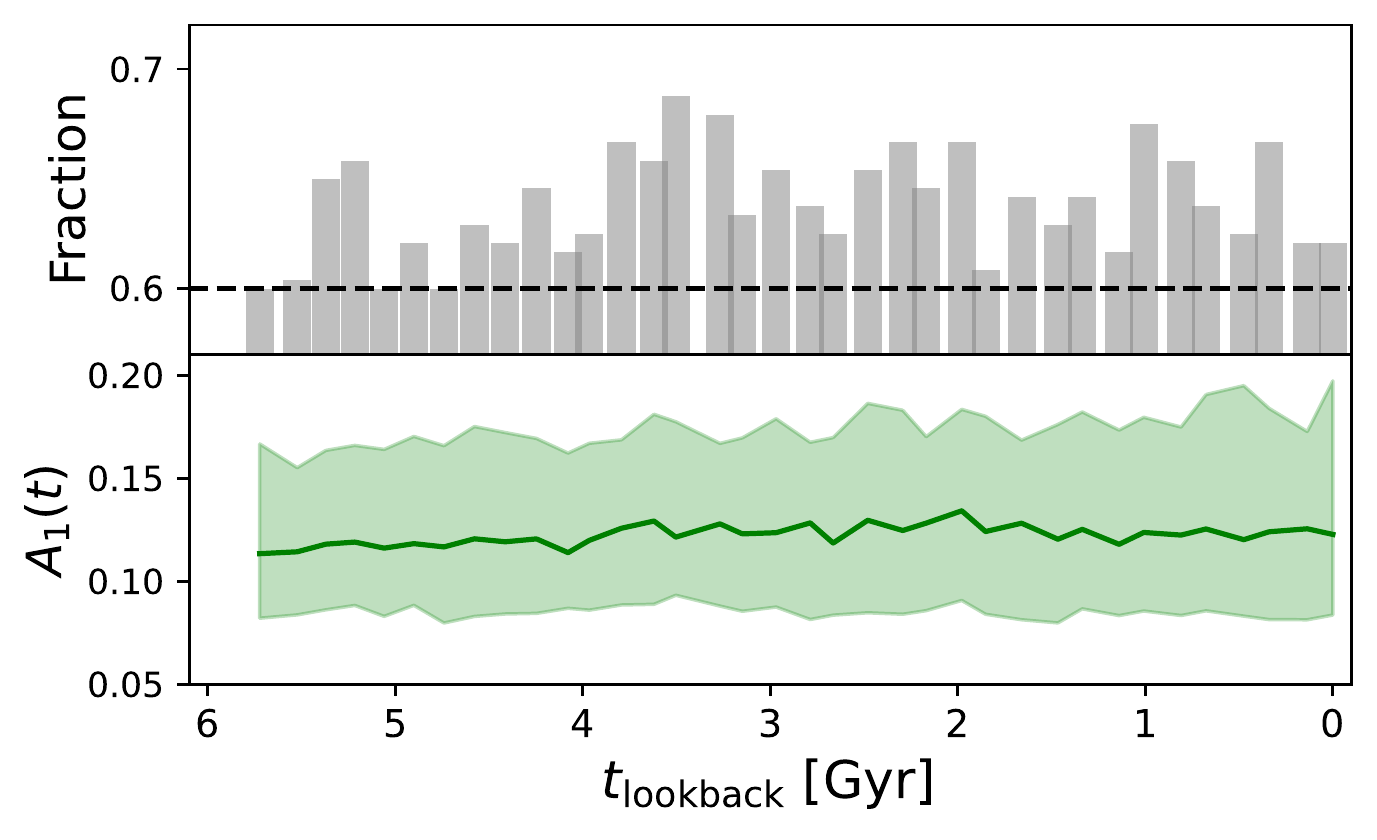}
\caption{ {\it Top panel:} Fraction of galaxies that show \Am$>0.1$ during the last 6 Gyr. {\it Bottom panel: }Distribution of $A_1$ in the sample in function of time. The green line correspond to the median of $A_1$ and the green region covers the 25$^{\rm th}$ to 75$^{\rm th}$ percentile of the $A_1$ distribution. We see that the medians of $A_1$ are around 0.125 during the last 6 Gyr. Our sample show that the fraction of galaxies with high lopsidedness are between 60 to 70 percent in that range of time}.
\label{fig:A1sampleintheTime}
\end{center}
\end{figure}

\subsubsection{Frequency of \Am}
\label{sec:freq_A1}

\begin{figure}
\begin{center}
\includegraphics[trim={0cm 0cm 2cm 1cm},clip,width=\linewidth]{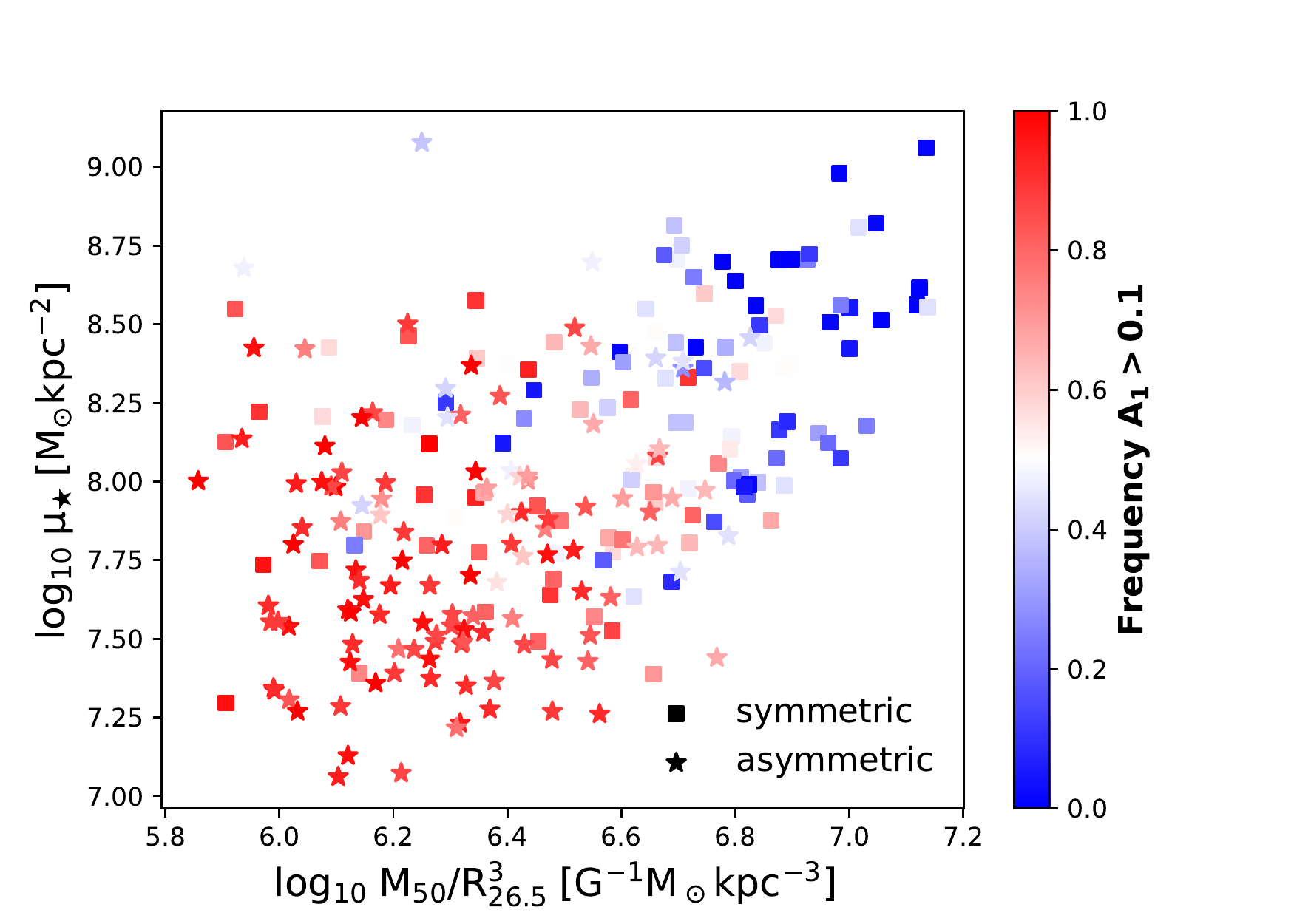}
\caption{Distribution of the central stellar density versus $\rm M_{50}/R^3_{26.5}$, which is a proxi of the tidal force exerted by the inner disc on  its outskirts. Galaxies are separated between symmetric (squared) and asymmetric (star), using \Am\ at $z=0$. The symbols are colour coded according to the fraction of time they experienced of significant lopsided perturbation ($A_1 > 0.1$) during the last  6 Gyr of evolution. Galaxies with lower $M_{50}/R^3_{26.5}$ tend to show lopsided distortions for longer periods. Interestingly, there are symmetric cases at the present-day that have spent long periods as lopsided (red squares). This galaxies typically show low $M_{50}/R^3_{26.5}$. Conversely, it is less common to see present-day lopsided galaxies with small $A_1 > 0.1$ frequency.
}\label{plotsMUTidalA1frac}
\end{center}
\end{figure}

As previously discussed, Fig. \ref{statesample_A1} shows that, at the present-day, a  62 percent of our simulated galactic discs are significantly lopsided (\Am$>0.1$). This suggest that this type of perturbations are either long-lived \citep[e.g.][]{Jog2009}, or short-lived but repeatedly  re-excited by subsequent perturbations \citep[see e.g.][]{Ghosh2022}. We explore this by following the time evolution of our simulated galaxies, and quantifying  the fraction of time they present a significant lopsided perturbation over the last 6 Gyr. In practice, we proceed as follow. We identify our galactic models in  the 36 snapshots available during the last 6 Gyr of evolution and compute, on each of them, the $A_1$ parameter. This parameter is calculated by fixing the value of $R_{26.5}$  at its present-day value. We have tested that our results do not significantly vary if we take into account the evolution of the optical radius. In Fig. \ref{fig:A1sampleintheTime}, we explore the distribution of \Am\ in our sample over the past 6 Gyr (bottom panel). The median of \Am (green line) is around 0.125 during this period. The green region cover the 25$^{\rm th}$ to 75$^{\rm th}$ percentiles of the \Am\ distribution of our sample, which does not exceed 0.2 for the 75$^{\rm th}$ percentile and does not fall below 0.075 for the 25$^{\rm th}$ percentile. In the top panel we show the fraction of galaxies that display a high amplitude of $m=1$ component (\Am$>0.1$), showing that around 60 to 70 percent of the galaxies in our sample exhibit high lopsidedness during this time range. That suggests that lopsided perturbation is a very frequent phenomenon for disc galaxies.

Fig.~\ref{plotsMUTidalA1frac} shows the distribution of galaxies in \mus~ versus $M_{50}/R^3_{26.5}$ space, colour coded according to the 
fraction of time each simulated galaxies experienced 
$A_1 > 0.1$  within the last 6 Gyr. Present-day lopsided galaxies, defined as in  Section \ref{Properties}, are shown with star symbols, whereas their symmetrical counterparts, with square symbols. In general, we find that symmetric galaxies (high \mus~ and $M_{50}/R^3_{26.5}$ values),  show low $A_1$ values throughout the latest 6 Gyr of evolution. In other words, strongly gravitationally cohesive galaxies have remained symmetric over most of the corresponding period of time (blue colors). On the other hand, we find that lopsided galaxies (typically weakly gravitationally cohesive) have remained lopsided ($A_1 > 0.1$) over a significant fraction of the latest 6 Gyr (red colors). There are however several examples of  galaxies that have been lopsided  over most of this period, but at the present day have a symmetric configuration (see red squares). Note as well that it is less common to find present-day lopsided galaxies with  low frequency of $A_1$. 

Our results suggest that lopsided perturbation are typically long-lived, rather than short-lived  but repeatedly re-excited. We further explore this in the following Section, where we follow the time evolution of a number representative galaxy models.

\subsection{Main driving agents}
\label{sec:driving}

As discussed in Section~\ref{sec:intro}, several different mechanism have been proposed as main driving agents for this type of morphological perturbation. The mechanisms range from direct tidal perturbations from relatively massive satellites, torques associated with perturbed underlying DM halos, and the non-axysimmetric accretion of cold gas, among others. In this Section we explore whether there is a dominant mechanism driving lopsidedness in our simulated galaxies.

\begin{figure*}
\begin{center}
\includegraphics[trim={0cm 3cm 1.3cm 1.5cm},clip,width=0.42\linewidth]{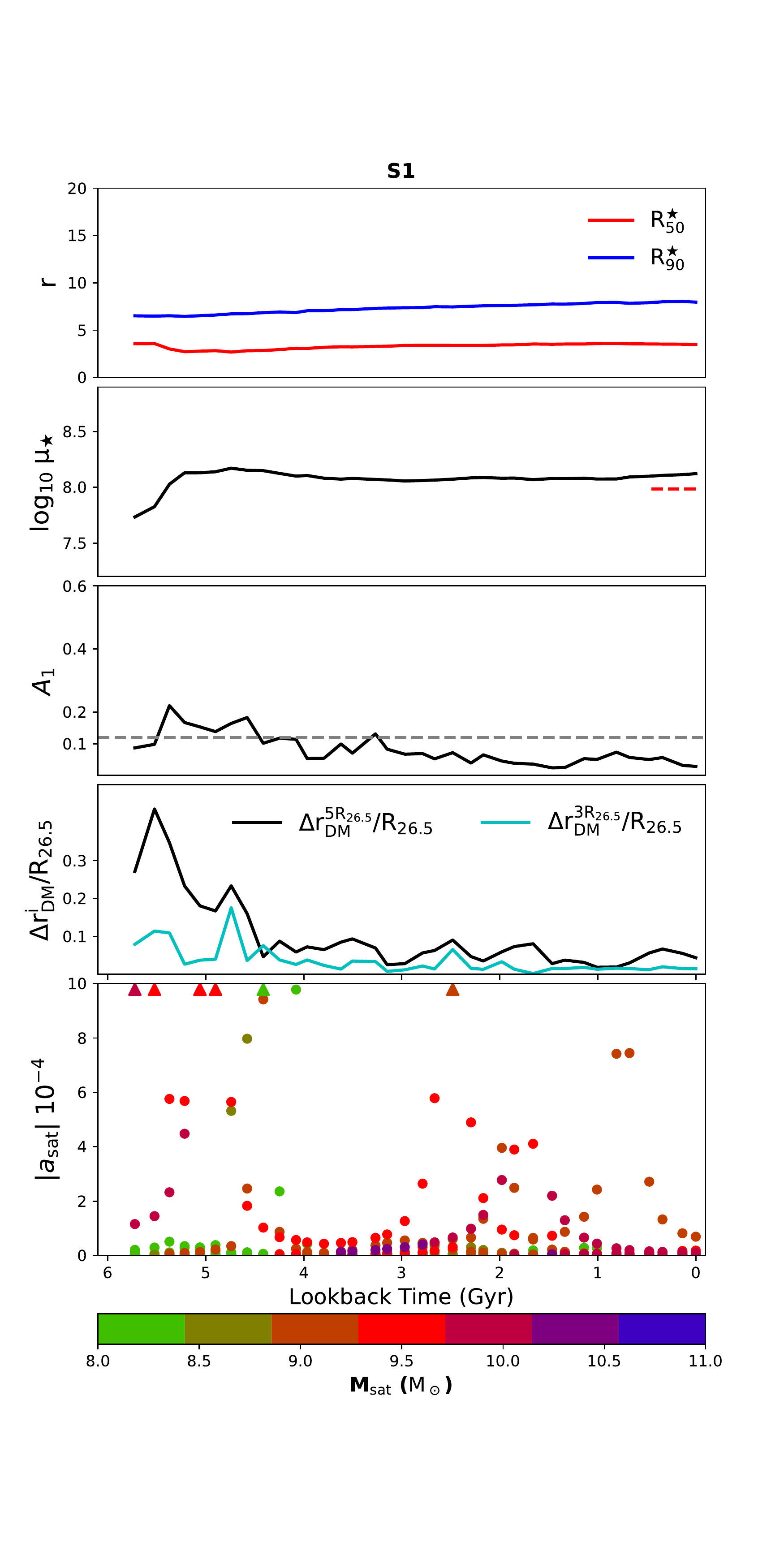}
\includegraphics[trim={0cm 3cm 1.3cm 1.5cm},clip,width=0.42\linewidth]{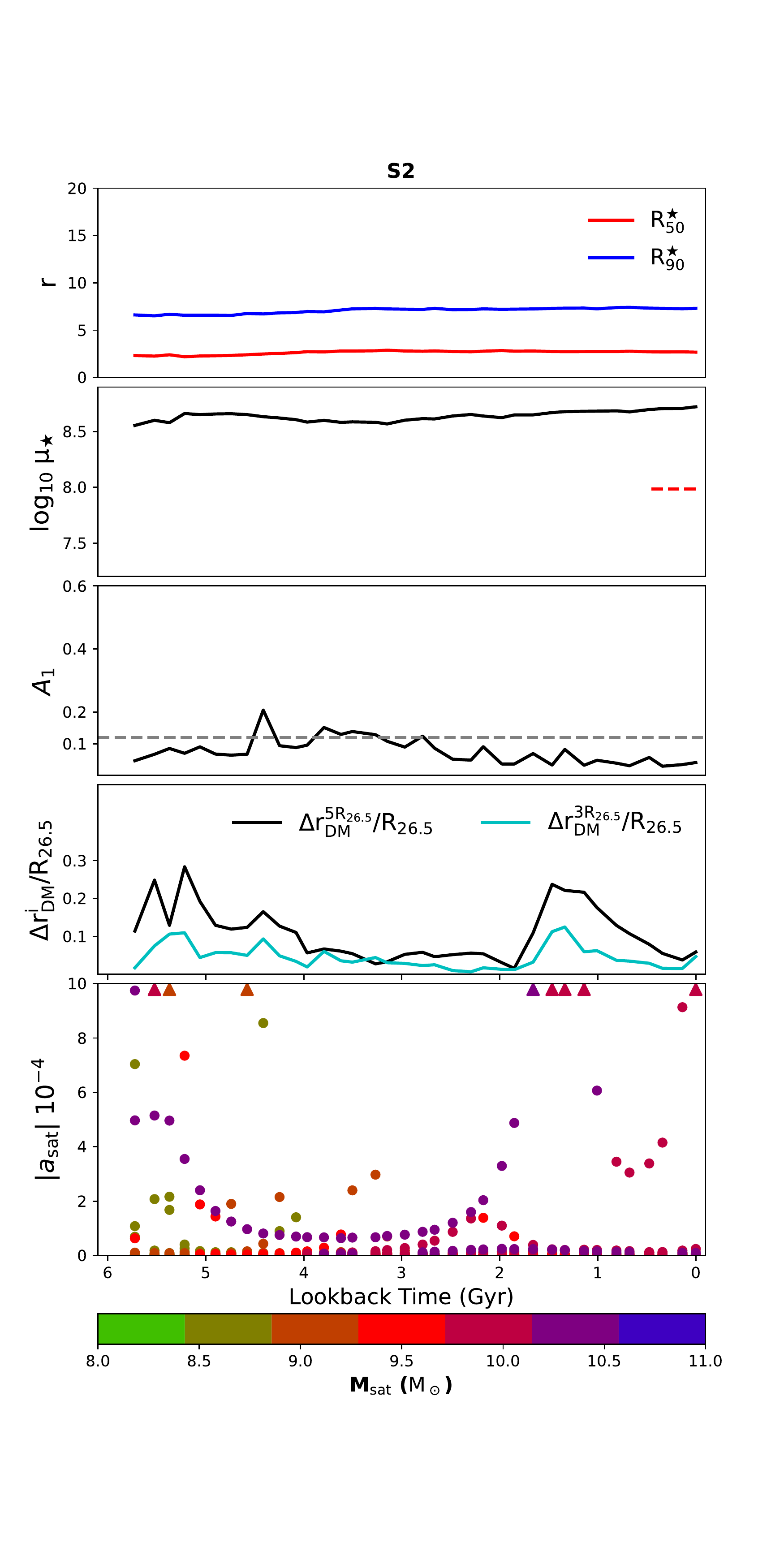}
\caption{Examples of two typical present-day symmetric galaxies, S1 and S2. {\it From top to bottom}, {\textit{First panel}}: Time evolution of the scale length parameters, \Rhalf\; and \Rt, during the last 6 Gyr of evolution. {\textit{Second panel}}: Evolution of the central surface density, \mus. The red dashed line on the right side of each panel shows the corresponding present-day   median of \mus. {\textit{Third panel}}: \Am\; as a function of time. The black dashed line correspond to the $\hat{A}_1$ threshold, extracted from the full sample distribution at $z=0$. {\textit{Fourth panel}}: Time evolution of the offset between the DM halo center of mass and its density cusp, $\Delta r^{i}_{\rm DM}$. We show the evolution of $\Delta r^{i}_{\rm DM}$ calculated within two spatial regions, 3\Rgal\; and 5\Rgal\; (see Sec. \ref{MeasureDMCM}). {\textit{Fifth panel}}: Time evolution of tidal field exerted on the host galaxy by its 10 most massive satellites.Triangles indicate tidal field values that are above the Y-axis limit. Symbols are colour coded as a function of the total mass of the corresponding satellite.
}\label{fig:plotsSymTime}
\end{center}
\end{figure*}

\begin{figure*}
\includegraphics[trim={0cm 3cm 1.3cm 1.5cm},clip,width=0.42\linewidth]{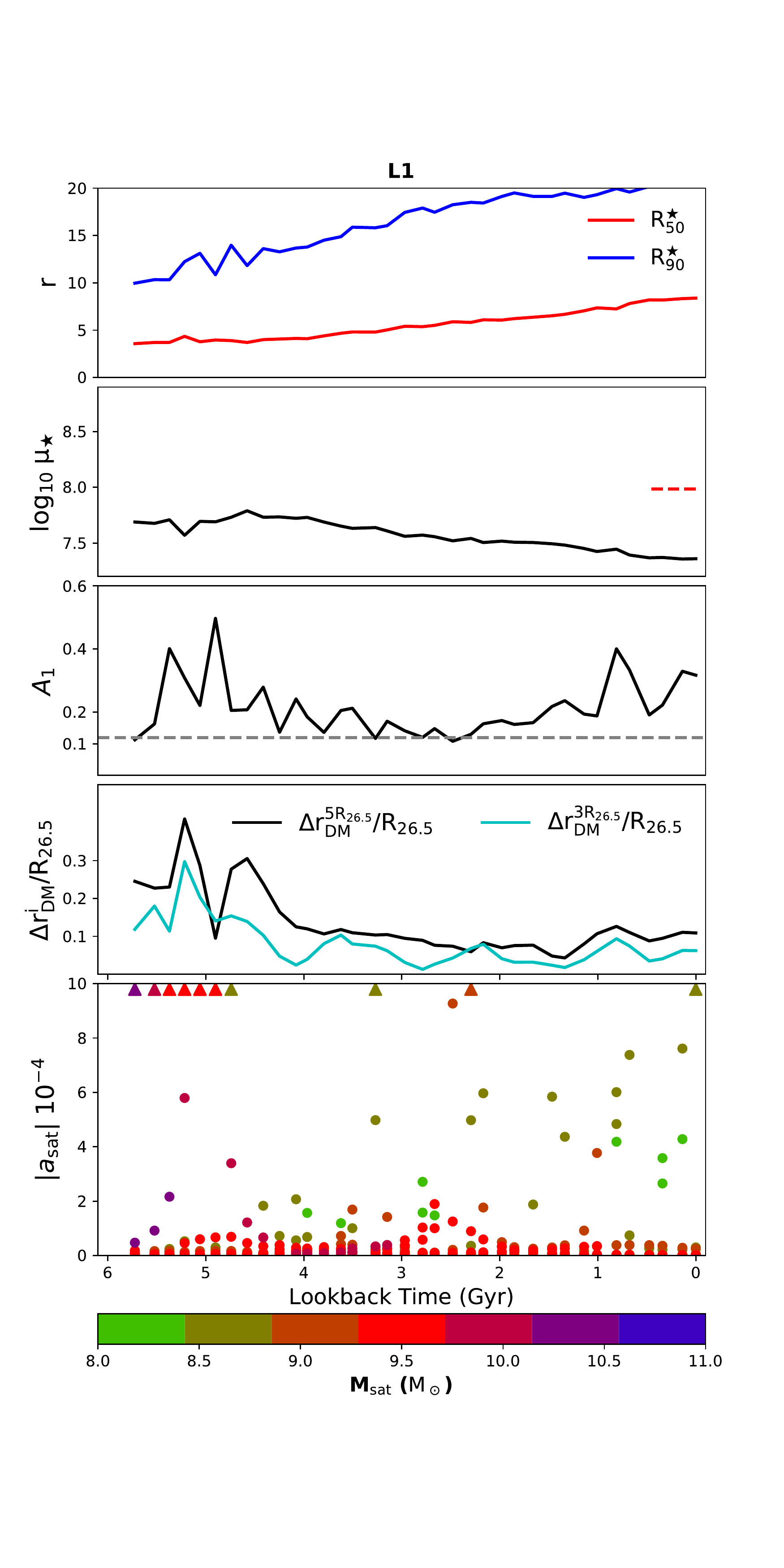}
\includegraphics[trim={0cm 3cm 1.3cm 1.5cm},clip,width=0.42\linewidth]{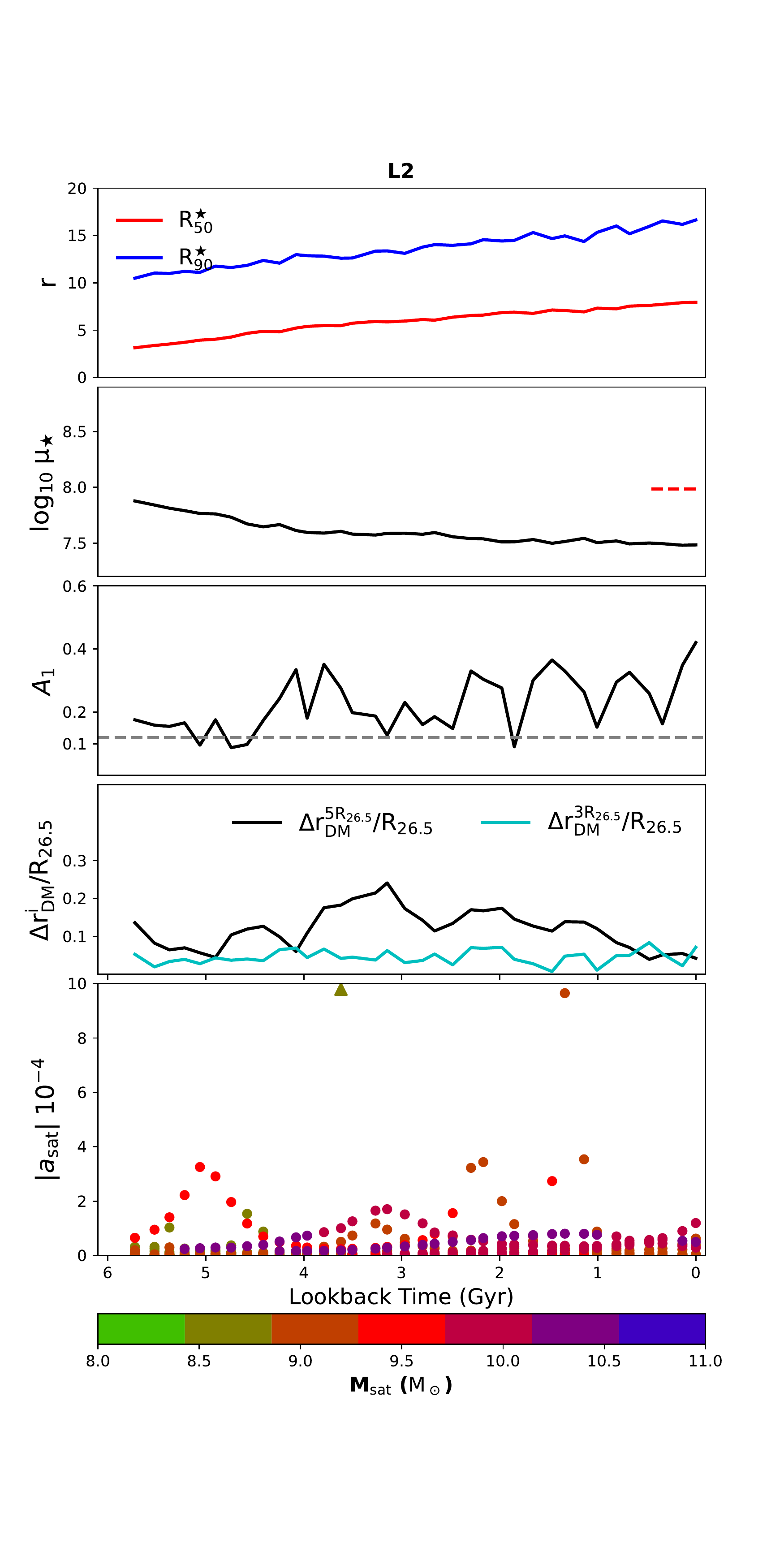}
\caption{As in Fig. \ref{fig:plotsSymTime}, for two typical present-day Lopsided galaxies, L1 and L2.}\label{fig:plotsAsymTime}
\end{figure*}

\begin{figure}
\begin{center}
\includegraphics[trim={0.5cm 0cm 2cm 0.5cm},clip,width=\linewidth]{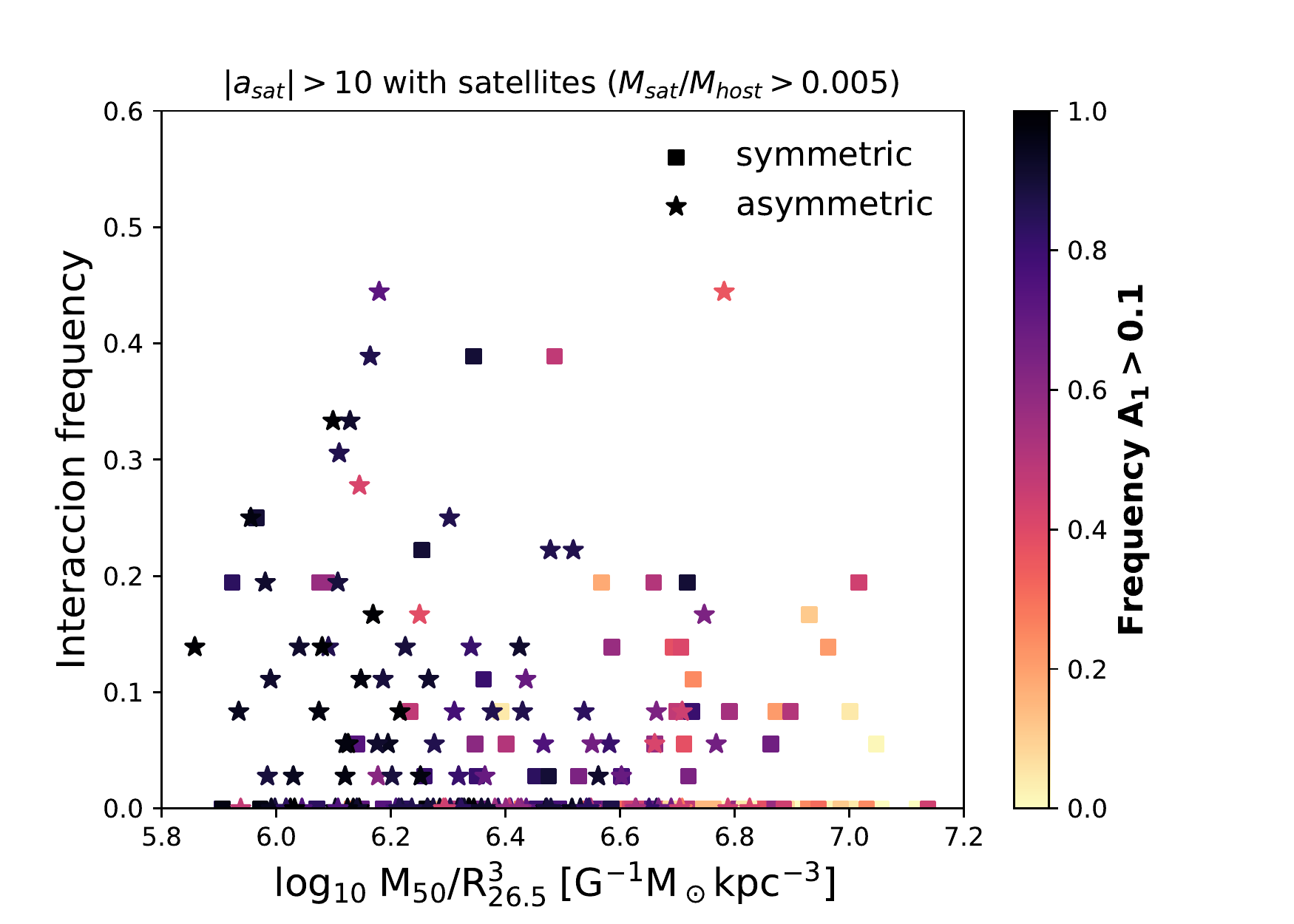}
\caption{Fraction of time a given host has strongly interacted with satellites of $M_{\rm sat} \geq 10^{9.5}$ M$_{\odot}$ during the last 6 Gyr of evolution in function of self-gravity proxy. The color coding indicated the fraction of time galaxies presented a strong lopsided perturbation ($A_{1} > 0.1$). The different symbols indicated whether galaxies are  symmetric or asymmetric at the present-day. Note that no significant correlation is found between the fraction of time galactic discs
display lopsided perturbations and the fraction of time they experienced significant satellite tidal interactions.}
\label{fig:intfrec}
\end{center}
\end{figure}

\subsubsection{Individual examples}
\label{sec:ind_sat}

Before analyzing the whole sample of galaxy models in a statistical manner, we first analyze in detail a couple of typical examples of present-day symmetric and lopsided galaxies.

We first focus on two  examples of typical present-day symmetric galaxies. The top panels of   Fig.~\ref{fig:plotsSymTime} show that, as discussed in Sec~\ref{sec:radiusintime}, symmetric galaxies typically  do not experience a substantial growth in size over the last 6 Gyr of evolution. Both discs show nearly constant \Rhalf~ and \Rt~ values over this period of time. In the second panel (top to bottom) we show the time evolution of \mus. As expected, both galaxies show \mus\ values larger than the $z=0$ median (red dashed line, $\hat{\mu}_\star=10^{7.98}$). In addition \mus\ show nearly constant values over this period of time. The large central surface density and small size render strong cohesiveness and thus resilience to perturbations. Indeed, as shown in the third panel, their $A_1$ value remains mainly below the $\hat{A}_1$ threshold, indicating that these galaxies have remained  symmetric over most of this period. We note, however, that the evolution of $A_1$ shows moderate increments over short spans of times. For example, for galaxy {\bf S1}, $A_1$ rises over $\hat{A}_1$ at a lookback time, $t_{\rm lb} \sim 5$ Gyr. To explore the origin of this short-lived lopsided mode  we quantify in the fourth and fifth panels  the interaction of this galactic disc with its environment. We first look at the time evolution of the  offset of the halo center of mass, CoM, with respect to is density cusp, $\Delta r^{i}_{\rm DM}$ (fourth panel).  Interestingly, $\Delta r^{i}_{\rm DM}$ peaks during the same period. This is noticeable  when considering DM particles up to a distance of $5R_{26.5}$. Note as well that this offset of the DM halo CoM is short lived and directly related to a strong tidal interaction with a massive satellite galaxy. This is shown on the bottom panel, where we show the time evolution of tidal field exerted on the host by its  10 most massive satellites as a function of time, i.e. $|a_{\rm sat}| = G M_{\rm sat}/R^3_{\rm sat}$. Here $M_{\rm sat}$ and $R_{\rm sat}$ are the total mass of the satellite and the distance between the satellite and its host galaxy. We notice that the galactic disc strongly interacts with a massive satellite ($M_{\rm sat} \sim 10^{10}$ M$_\odot$) at  $t_{\rm lb} \sim 5$ Gyr. This strong interaction is behind the brief distortion of the host outer DM halo, and the temporary onset of a mild $A_1$ perturbation. For this large \mus~ simulated galaxy, and in agreement with \citet{Ghosh2022}, the perturbation rapidly dissipates and the amplitude of the disc $m=1$ mode remains below $\hat{A}_1$ for the rest of the period, even though a second significant interactions takes place later on ($t_{\rm lb} \sim 2.5$ Gyr).  

In the right panels  of Fig.~\ref{fig:plotsSymTime} we analyze a second example, S2, of a present-day symmetric galaxy. As before, the galaxy shows small  and nearly constant scale lengths over the whole the last 6 Gyr of evolution. It shows as well a nearly constant \mus~ value, but with a value larger than in the previous example, S1. The $A_1$ parameter typically remains below $\hat{A}_1$, except for short periods where it slightly raises over this threshold. When inspecting  interactions with its environment, we observe that S2 experienced two very strong interactions with a satellite galaxy of $M_{\rm tot} \approx 10^{10.5}$ M${\odot}$. These interactions take place at $t_{\rm lb} \approx 5.5$ and 1.5 Gyr (fifth panel), and both resulted in significant perturbations of the host DM halo (fourth panel). Yet, due to the large \mus, no associated response is observed in the evolution of the $A_1$ parameter.

 In Fig.~\ref{fig:plotsAsymTime} we now explore two examples of strongly lopsided galaxies, L1 and L2. The top panels show that, contrary to the symmetric  cases, these galaxies experienced a consistent growth in size over the latest 6 Gyr, which resulted in a decrement of their \mus. As a result, the central surface density of these simulated galaxies is significantly lower (second panels) than in the symmetric examples. The third panels show that, in both cases, $A_1$ has mainly remained above our threshold,  $\hat{A}_1$, indicating long lived lopsided modes. In particular, for L1 (left panels), we find that the galaxy experienced a relatively  strong tidal interaction with a satellite of $M_{\rm sat} \sim 10^{10.5}$ M$_{\odot}$ at $t_{\rm lb} \approx 5$ Gyr. Due to the low values of \mus, and contrary to the S1 case, this interaction excited a strong lopsided mode as well as a shift of the DM CoM with respect to its density cusp. The lopsided perturbation slightly wanes over time, but it always remains over $\hat{A}_1$. At $t_{\rm lb} \approx 2$ Gyr the disc experienced a second significant tidal interaction ($M_{\rm sat} \sim 10^{10}$ M$_{\odot}$) that enhances the lopsided perturbation, raising the value of $A_1 \approx 0.35$ until the present-day. On the other hand even though L2 (right panels) shows a value of $A_1 > \hat{A}_1$ over most of the last 6 Gyr, it did not experience any significant interaction with  massive satellites $(M_{\rm sat} > 10^{10}$ M$_{\odot})$. Interestingly, the CoM of outer DM halo shows a significant shift with respect to its cusp during this period, with values $\Delta r_{\rm DM}^{5R_{26.5}}$ as large as 20 percent of $R_{26.5}$. Such perturbed DM halo could be behind the long lived lopsided perturbation in this galactic disc \citep[see e.g.][]{Jog2009}. We will explore in detail this particular kind of models in a follow up study.

\subsubsection{Statistical characterization of the impact of interactions}

In the previous Section we discussed two examples of stellar discs that interacted with their nearby environment and developed very strong lopsided perturbations. On the other hand, we also discussed examples of galaxies that strongly interacted with their environments but did not develop significant lopsided modes on their  discs. The main difference between these two sets of objects is their central surface density, \mus, which set the gravitational cohesiveness of the disc outskirts. In this Section we explore what are the main agents driving lopsided perturbations in low \mus~galaxies.

We start by quantifying significant tidal interaction with satellite galaxies within the last 6 Gyr of evolution. As in Section \ref{sec:ind_sat}, for each galaxy in our sample, we compute $|a_{\rm sat}|$ as function of lookback time. Based on Figures  \ref{fig:plotsSymTime} and \ref{fig:plotsAsymTime}, and the results shown in \citet{Gomez2017}, we first quantify the fraction of time galactic discs were exposed to   $|a_{\rm sat}| > 10$ from satellites with mass ratios $M_{\rm sat}/M_{\rm host} > 0.005$. Lower values of $|a_{\rm sat}|$ do not yield to global perturbations in the discs. For a MW-mass host, the chosen threshold in the mass ratio allows interaction with satellites of $M_{\rm sat} \geq 10^{9.5}$ M$_{\odot}$. Lower mass satellite are unlikely to induced significant global vertical perturbations \citep{Gomez2017}. Figure \ref{fig:intfrec} shows our sample of galactic discs in satellite interaction frequency versus $M_{50}/R^3_{26.5}$ space. Points are color coded by the fraction of time each disc presents a significant lopsided perturbation (see Sec.\ref{sec:freq_A1}). As before, we notice that galaxies with lower frequencies of $A_1 > 0.1$ (light colors) tend to have lower values of $M_{50}/R^3_{26.5}$. Interestingly we find no significant correlation between the fraction of time galactic discs display lopsided perturbations and the fraction of time they experienced significant satellite tidal interactions. In particular, a significant number of simulated disc galaxies (35 per cent of the full sample) did not experience significant interactions during the last 6 Gyr, but  nevertheless, have a long-lasting lopsided perturbation over most of that period. This supports our conclusion that direct tidal interaction with satellites galaxies is just one plausible channel for inducing lopsided perturbation, and not the main driving agent.
Our results are in agreement with those presented by \citet{Bournaud2005} who shows with a sample of 149 observed galaxies that the $m=1$ amplitude is uncorrelated with the presence of companions.

In addition to  direct tidal torques exerted by satellites, galactic discs can respond to the gravitational interaction with a distorted DM halo \citep{FGomez2015,GaravitoCamargo2019,Laporte2018a}. One of the first attempts to statistically study asymmetries in the inner regions of dark haloes, and their possible relation to the accretion of external material on to these regions, was provided by \citet[][hereafter GW06]{gao}. Based on the large statistic provided by the Millennium Simulation \citep{Springel2005}, they studied asymmetries in the density distribution of DM halos, selected with masses ranging from MW-mass to cluster mass hosts. They showed that such asymmetries are not uncommon, and that the frequency with which they arise depends on host mass. While 20 per cent of cluster haloes have density centres that are separated from barycentre by more than 20 per cent of the virial radius, only seven per cent of MW haloes have such large asymmetries. 

Following GW06, we examine the distribution of the offsets of central DM halos extracted from the TNG50-1  simulation and its  DM only simulation counterpart. Our goal is to test whether we recover the results presented in GW06, based purely on DM only models, and test whether the addition of baryons has an effect on this statistics. Since our work in centered around late type galaxies, we focus on three sets of models selected by halo mass, \Mcrit. The less massive set contains haloes with \Mcrit~ between $\rm 10^{11}$ to $\rm 5x10^{11}\; M_{\odot}$; the intermediate set between $\rm 5x10^{11}$ to $\rm 10^{12}\; M_\odot$, and the more massive between $\rm 10^{12}$ to $\rm 5x10^{12}\; M_\odot$. In the table \ref{tableGao} we show the results of this selection for each simulation.

\begin{figure}
\begin{center}

\includegraphics[trim={0cm 1cm 1.2cm 2cm},clip,width=\linewidth]{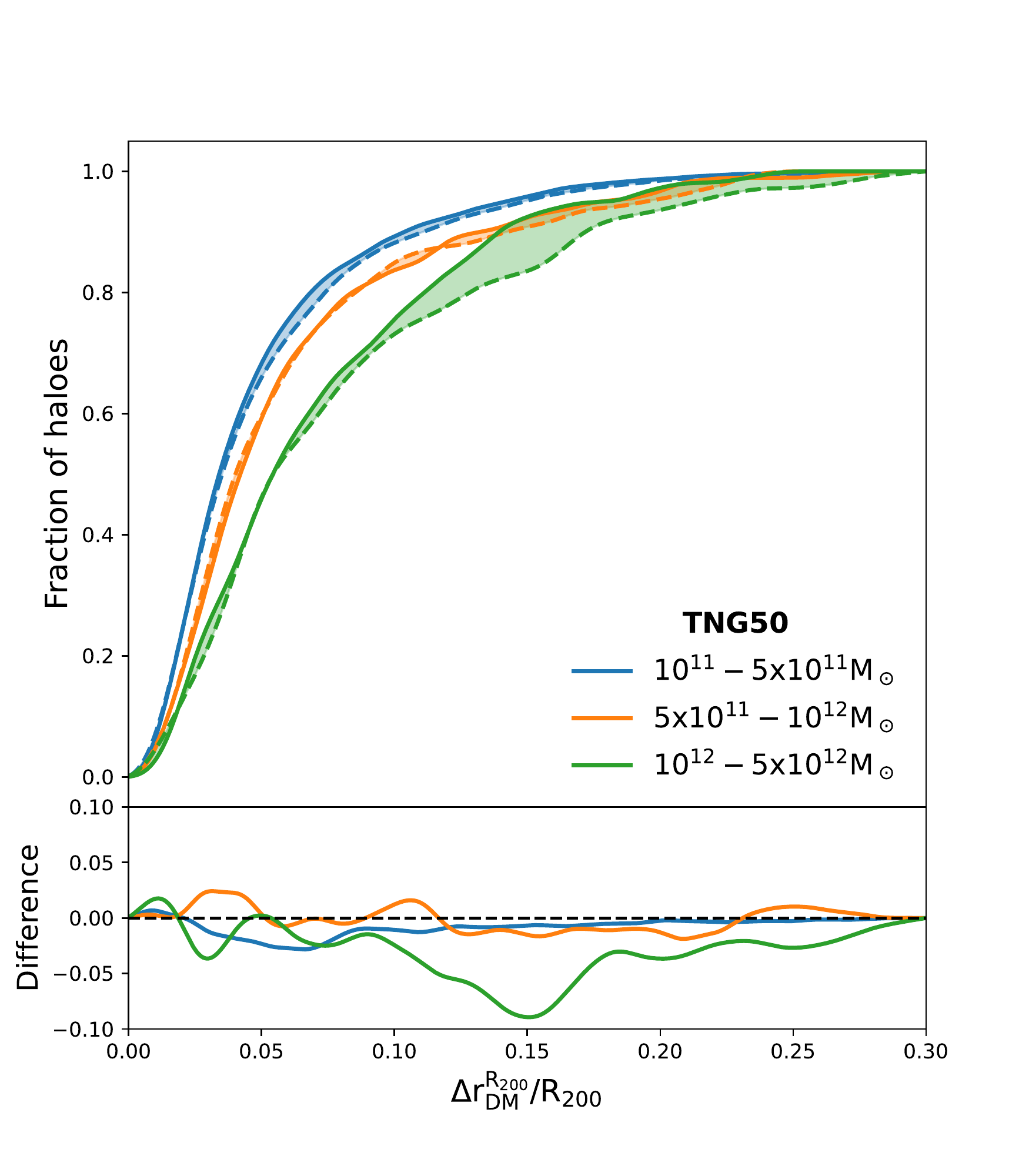}

\caption{Top panel: Cumulative distribution function of offset between the halo center of mass and their density cusp, \deltar. The  colour coded lines show the results obtained from different halo mass ranges.  For this calculation DM particles within each halo's $R_{200}$ are considered. Solid and dashed lines show the results obtained from the full hydrodynamical and the DM only simulations, respectively. Bottom panel: differences between cumulative distribution function obtain from the full hydrodynamical and the DM only simulations.}\label{plotsgao}
\end{center}
\end{figure}

For each simulated galaxy, we compute   $\Delta r^{\rm R_{200}}_{\rm DM}$~(see Eq.\ref{eq:delta_rdm}) at the present-day considering DM particles within \Rcrit. In the top panel of Figure \ref{plotsgao}, we show the $\Delta r^{\rm R_{200}}_{\rm DM}$~  cumulative distribution function (CDF) for the three halo subsets. The solid and dashed lines show the results obtained from the full hydrodynamical simulation and the DM only, respectively. To facilitate the comparison between these different haloes, each \deltar~ has been normalized by the corresponding \Rcrit. Our results based on the DM only simulations are in good agreement GW06. We find that more massive haloes tend to have larger asymmetries. Indeed, while $\sim 8$ percent of halos with $10^{12} < M_{200} < 5\times10^{12}$ M$_{\odot}$ show $\Delta r^{\rm R_{200}}_{\rm DM}$~$ > 20$ percent, for haloes with $10^{11} < M_{200} < 5\times10^{11}$ M$_{\odot}$ only $\sim 1.5$ percent show such large asymmetries. Comparison with the results obtained with the full-physics models shows that these trends are not significantly affected by the addition of baryons. The shaded areas highlight the differences between the DM only and the hydrodynamical simulations. Note that only the larger mass halo subset shows a slightly larger fraction of halos with $\Delta r^{\rm R_{200}}_{\rm DM}$~$\gtrsim 0.15$ in the hydrodynamical simulation. However, this difference mainly arise from the low number statistics associated to mass bin\footnote{We have confirmed that his difference is due to low number statistics by repeating the analysis on the larger volume simulations TNG100-1.}. The similarities between both simulations are better highlighted in the bottom panel of Fig.~\ref{plotsgao}, where we show the difference between both CDFs. 

\begin{figure}
\begin{center}
\includegraphics[trim={0cm 0cm 1.2cm 1cm},clip,width=\linewidth]{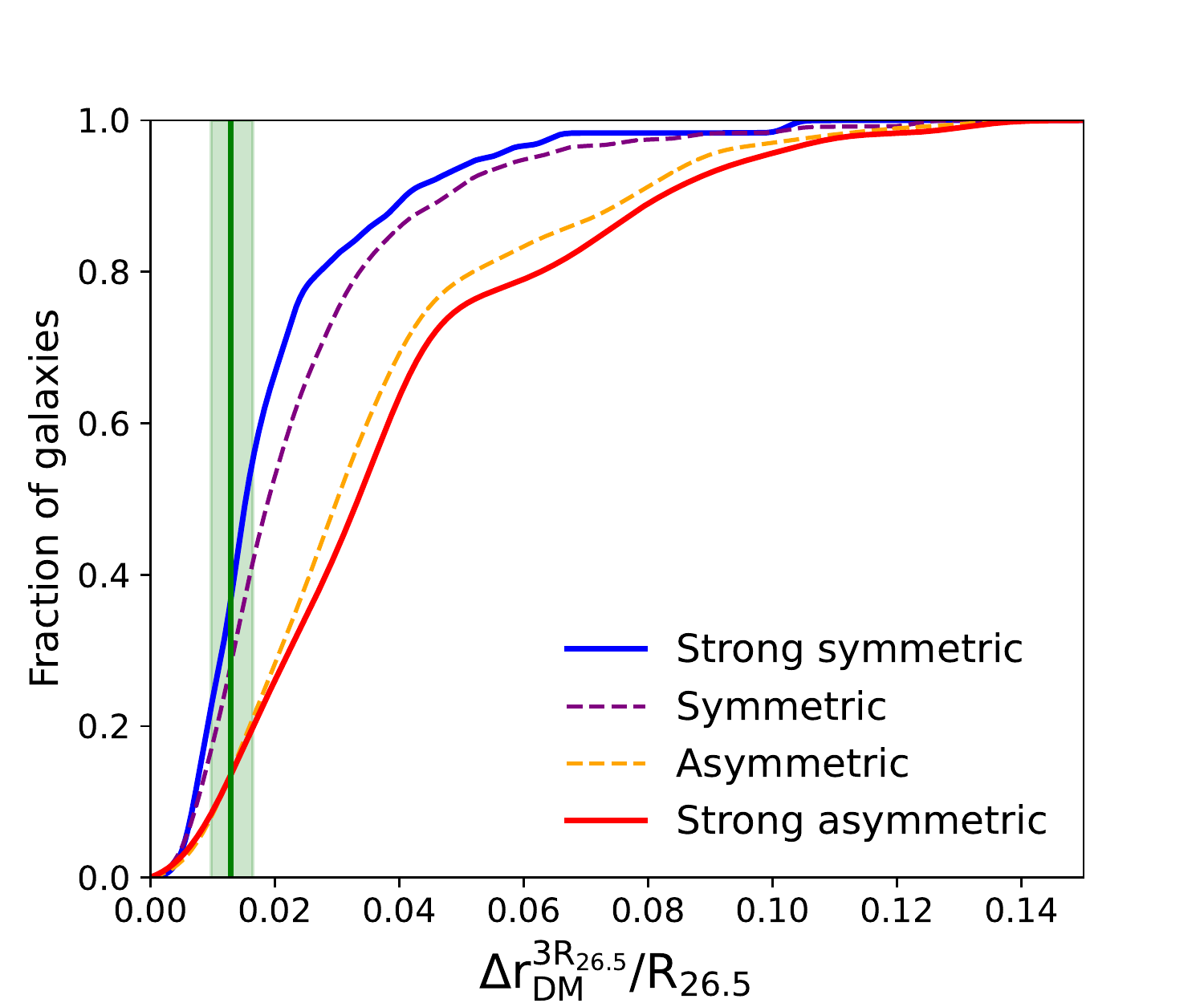}

\caption{ Cumulative distribution function of the offset between the halo center of mass and their density cusp,  $\Delta r_{\rm DM}^{3R_{26.5}}$. For this calculation DM particles within each galaxy $3\times$\Rgal\ are considered.  The dashed purple (solid bule) and orange (red) lines show the results obtain from the (strong) symmetric and asymmetric subsamples, respectively. The vertical green line indicates the median of the distribution obtained after normalizing the  gravitational softening, $\epsilon_{\rm DM}$, by the \Rgal\ of each galaxy. The shaded green area encloses the 25th and 75th percentile of the corresponding distribution.}\label{GTCDFDM}
\end{center}
\end{figure}

As previously discussed in GW06, these DM halo asymmetries could be related to visible asymmetric phenomena in galaxies, among them lopsidedness. To explore this, we show  the $\Delta r^{\rm i}_{\rm DM}$~ CDF, now considering only galaxies selected by the criteria defined in Section \ref{sec:crit}. For this analysis we focus on perturbations within the inner DM halo,  $3 \times$\Rgal\; (Figure \ref{GTCDFDM}), since this is the region that can exert significant torque on the embedded discs \citep[e.g][]{Gomez2016}. We first explore the subset of galaxies that are strongly symmetric and strongly lopsided. The corresponding CDFs are shown in solid red and blue lines respectively.  The vertical green line indicates the  mean of the gravitational softening length, $\epsilon_{\rm DM}$ (see table \ref{tablesimu}), obtained after normalizing $\epsilon_{\rm DM}$ by the \Rgal~ of each galaxy. The shaded region covers 25th and 75th percentiles of this distribution. The figure clearly shows that symmetric galaxies tend to have small $\Delta r^{\rm 3R_{26.5}}_{\rm DM}$, indicating very similar spatial location for the DM CoM of the center of density. Only 5 percent of the symmetric galaxies show values of $\Delta r^{\rm 3R_{26.5}}_{\rm DM}$~ $> 0.05$. The CDF for the asymmetric galaxies shows a different behaviour. It is clear that asymmetric galaxies tend to show significantly large $\Delta r^{\rm 3R_{26.5}}_{\rm DM}$~ than their symmetric counterparts. Indeed, $\approx 30$ percent of the disc galaxies shows $\Delta r^{\rm 3R_{26.5}}_{\rm DM}$~ $> 0.05$. Yet, as discussed in Section \ref{sec:ind_sat}, we find a large number of lopsided galaxies show very small $\Delta r^{\rm 3R_{26.5}}_{\rm DM}$, indicating that this is not necessarily the main driver behind their perturbations.

\begin{table}
\begin{center}
\caption{Number of haloes for each \Mcrit~ range between the two homologous runs; baryonic + DM simulation and only DM.}
\label{tableGao}
\begin{tabular}{lrr}
\hline
      & \multicolumn{2}{c}{\bf TNG50-1}                   \\
        & Baryon+DM             & DM only               \\
\hline
$\rm 10^{11}$ - $\rm 5x10^{11}\; M_{\odot}$ & 1251                  & 1352                  \\
$\rm 5x10^{11}$ - $\rm 10^{12}\; M_\odot$   & 190                   & 185                   \\
$\rm 10^{12}$ - $\rm 5x10^{12}\; M_\odot$   & 168                   & 172                   \\

\hline

\end{tabular}
\end{center}
\end{table}

\section{Conclusions and discussion}
\label{sec:conclu}

In this paper we have studied disc galaxies that display a global $m=1$ non-axisymmetric  perturbation in their stellar mass distribution, more commonly known as a lopsided perturbation. We focused the analysis on a sample of MW-mass like galaxies from the fully cosmological hydrodynamical  simulation, TNG50 from IllustrisTNG project. Our sample was built selecting central subhalos with \Mcrit\ within the range $10^{11.5}$ to  $10^{12.5}~M_{\odot}$. To consider well-resolved disc-dominated galaxies we imposed a threshold in the D/T ratio of 0.5, and only selected galaxies with more than $10^4$ stellar particles within a subhalo. From this criteria, 240 late-type galaxies with total stellar mass between $\rm 10^{9.5}$ and $\rm 10^{11.2}\;M_{\odot}$ were selected. Lopsidedness in the discs were quantified by computing the amplitude of the $m=1$ Fourier mode of the stellar density distribution, \Am. Based on this parameter,  we classified our galaxies as symmetrical and asymmetrical (i.e. lopsided) cases. 

We find that in our simulated galaxy sample the main characteristics of such lopsided perturbations are in good agreement with observations. In lopsided galaxies, the radial profile of the $m=1$  mode amplitude,  \Am$(R)$, increases with radius in the outer disc regions, while in the inner parts it remains flat and close to zero. The radius at which the transition takes place is $\approx 0.5$\Rgal, in agreement with previous observational works \citep[e.g.][]{Rix1995,Bournaud2005}. Furthermore, lopsided galaxies exhibit a nearly constant or midly varying radial distribution of phase angles, indicating a slow winding of the phase angle in the outer disc \citep{Saha2007,Ghosh2022}. Based on this, we computed for each  simulated galaxy a characteristic \Am\ value, which corresponds to the average of \Am$(R)$ between $0.5R_{26.5} < R < 1.1R_{26.5}$. We find that the distribution of this characteristic \Am\ parameter is also in good agreement with observations, that measured in large observational samples the \Am\ distribution, considering similar galactic regions \citep{Bournaud2005,Reichard2008}. To highlight differences between lopsided and symmetrical galaxies, we focus on the analysis on the first and fourth quartiles of the \Am\;distribution. We call these subsets strong symmetric and strong asymmetric galaxies, respectively.

When analyzing the present-day structural parameters of our sample, we find that lopsided galaxies tend to be more disc-dominated than they symmetrical counterparts. This trend suggests that the presence of centrally pressure supported component plays an important role on setting the lopsidedness strength. This is in agreement with previous works, which found that  the fraction of lopsided galaxies increases with galaxy Hubble type, being late-type galaxies the population with the highest fraction \citep{Rix1995,Bournaud2005,Conselice2000}. Following \citetalias{Reichard2008}, we also characterized our sample through the following present-day structural parameters: stellar half-mass radius, \Rhalf,  total stellar mass, $\rm M^{\star}$,  central surface density, \mus\, and  stellar concentration, $\rm C_{\star}$. Focusing on the strong subsamples, we find that both lopsided and symmetric galaxies show very similar $\rm C_{\star}$ distributions. We also find that strongly lopsided galaxies tend to have more extended central regions and to be slightly less massive than their counterpart symmetrical. However, the most  strong (anti)correlation we find is between \mus\ and \Am. Indeed, the strong subsamples show very different distributions of \mus, with lopsided disc systematically showing lower \mus\ values.  These results are consistent with the observational findings from \citetalias{Reichard2008}, suggesting that galaxies with lower central density could be more susceptible to different types of interaction and, thus, more prone to the excitation of a lopsided modes. Based on these results, we show that what regulates whether a galaxy develops strong lopsided modes is the self-gravitating nature of the inner galactic regions. Discs with denser inner regions are more gravitationally cohesive and thus, less  prone to develop lopsided perturbations in their external regions. Hence, our results hint toward a population of galaxies susceptible to lopsided perturbations, and not to a particular external driving source.

We have explored the time evolution of the main structural parameters that differentiate symmetric and lopsided galaxies, as well as the time evolution of the amplitude of the lopsided modes. We focus on their behaviour  during the last 6 Gyr of evolution. We observed that the percentage of galaxies in our sample with \Am$>0.1$ are between 60 and 70 during this range of time. Interestingly, we find that, while for symmetric galaxies \mus\ remains nearly constant through time, a
significant decay of \mus\ is observed in  lopsided galaxies. The main reason for this is the faster growth of the half-mass radius, \Rhalf, displayed by lopsided galaxies with respect to their symmetric counterparts. While both galaxies experienced similar growth rates of their \Mhalf, lopsided galaxies grow faster in size thus reducing their inner self-gravitational cohesion. Following \citet{grand2016}, we analyzed whether the halo spin, $\lambda$, is behind this faster growth rate of \Rhalf\ in lopsided objects. Interestingly, we find that galaxies with higher present-day $\lambda$ are typically less cohesive and show  higher values of $A_1$. On the other hand, galaxies with low $\lambda$ values are dominated by strongly self-gravitating discs and, thus, low $A_1$ values. Our results highlight an interesting morphology--halo connection for late type galaxies. 

We have also analyzed the main agents driving these perturbations. In agreement with previous studies, we have shown that  satellite interactions can excite lopsided modes \citep{Weinberg1995,Zaritsky_1997,Bournaud2005}. However, we find that up to $\sim 35$ per cent of the sample galaxy shows significant lopsided perturbations but, during the last 6 Gyr of evolution, did not experienced interactions with any satellite of mass ratio $M_{\rm sat}/M_{\rm host} >0.005$. Interestingly those galaxies present low values of \mus. This supports the conclusion that direct tidal interaction with satellite is a possible channel for inducing lopsided perturbation, but not the main driving agent. Our results are in agreement with those presented by \citet{Bournaud2005} who shows with a sample of 149 observed galaxies that the $m=1$ amplitude is uncorrelated with the presence of companions. 

Several studies have shown that galactic discs can also respond to tidal torques exerted by global perturbations of the host DM halo density distribution \citep{Weinberg1998,Gomez2016,Laporte2018a,Laporte2018,Hunt2021,Grand2022} . To examine whether this mechanism is an important driving agent of lopsided perturbations in our simulations, we quantified the distribution of  offsets between the CoM DM halo and the density cusp, $r_{\rm cusp}$ , of our halos, $\Delta r^{\rm R_{200}}_{\rm DM}$.  Previous studies based on the dark matter-only Millenium simulations \citep{gao} found that significant  distortions in the DM halos are not uncommon, and that the frequency with which they arise depends on host mass. Our analysis, based on simulations that incorporate a self-consistent treatment for the evolution of baryons, yielded very similar results. While $\sim 8$ percent of halos with $10^{12} < M_{200} < 5\times10^{12}$ M$_{\odot}$ show $\Delta r^{\rm R_{200}}_{\rm DM}$~$ > 20$ percent , only $\sim 1.5$ percent of halos with $10^{11} < M_{200} < 5\times10^{11}$ M$_{\odot}$  show such large asymmetries. Given this result, we studied whether halos with large offsets typically host lopsided galactic discs. Interestingly, we find that symmetric galaxies tend to have smaller distortions in their inner DM halos (within $3\times R_{26.5}$) than their lopsided counterparts. While only  five percent of the symmetric galaxies show values of $\Delta r^{\rm 3R_{26.5}}_{\rm DM}$~ $> 0.05$, $\approx 30$ percent of the lopsided galaxies do so. This results place torques from DM halo overdensity wake as another important mechanism behind the excitation of lopsided modes in galaxies with low central surface densities. In a follow up study we will quantify such torques  by decomposing the density and potential distributions using basis function expansions \citep[BFEs][]{GaravitoCamargo2021a,Cunningham2020,Johnson2023,Lilleengen2023}. Furthermore, we found that lopsided galaxies tend to live in high spin dark matter halos. Using the same simulation suite  \citet{Grand2017} showed that the present-day size of a stellar disc  is strongly related to the the spin of its halo. High spin halos tend to host extended galaxies with lower central surface densities, thus prone to develop lopsided perturbations. This result, together with  
the lopsided response of discs to overdensity wakes, indicates a new direction for understanding the halo-galaxy connection in lopsided galaxies.

In this work, we have shown that lopsidedness is a very frequent phenomenon in the history of galaxies. The discs of these galaxies are extended with low central surface densities. Their self-gravity makes them cohesively weak, and therefore easily susceptible to any type of interaction such as tidal torques exerted by distorted DM halos and minor mergers. Such galaxies tend to reside in high-spin and often highly asymmetric DM halos, revealing a connection between the halos and lopsided discs.

\section*{Acknowledgements}
SVL acknowledges financial support from ANID/"Beca de Doctorado Nacional"/21221776. FAG and SVL acknowledges financial support from CONICYT through the project FONDECYT Regular Nr. 1211370. FAG and SVL acknowledge funding from the Max Planck Society through a Partner Group grant. PBT acknowledges partial support from Fondecyt 1200703/2020 (ANID),  Nucleus Millennium ERIS ANID NCN2021-017. FAG, PBT and SVL acknowledge support from ANID BASAL project FB210003. CL acknowledges funding from the European Research Council (ERC) under the European Union’s Horizon 2020 research and innovation programme (grant agreement No. 852839).

\section*{Data availability}
The data used in this work is accessible via the IllustrisTNG public database\footnote{https://www.tng-project.org/data}.

\bibliographystyle{mnras}
\bibliography{tng}








\bsp	
\label{lastpage}
\end{document}